\documentclass[11pt,epsf]{article}
 \usepackage{amsmath}
 \usepackage{graphicx}
 \usepackage[merge,numbers,compress]{natbib}
 \usepackage[T1]{fontenc}
 \usepackage{booktabs}
 \usepackage[dvipsnames,table]{xcolor}
 \usepackage{xspace}
 \usepackage{dcolumn}
 \usepackage{placeins}
 \usepackage{amsfonts}
 \usepackage{url}
 \usepackage[colorlinks=true,citecolor=blue!50!black,linkcolor=black]{hyperref}
 \usepackage{caption}
 \usepackage
 [subrefformat=parens,position=top,skip=-15pt,margin=15pt,justification=justified,singlelinecheck=false]
 {subcaption}

\newcommand{\R}[0]{\mathbb{R}}

\newcommand{\change}[1]{{{\color{black}{#1}}}}

\newcommand{\NN}{NN\xspace}

\def\citere#1{\mbox{Ref.~\cite{#1}}}

\newcommand{\rL}{\mathrm{L}}

\newcommand{\rd}{\mathrm d}
\newcommand{\nnb}{\nonumber}

\newcommand{\ie}{\emph{i.e.}\ }

\def\be{\begin{equation}}
\def\ee{\end{equation}}

\newcommand{\Pj}{\ensuremath{\text{j}}\xspace}
\newcommand{\Pp}{\ensuremath{\text{p}}\xspace}

\newcommand{\PW}{\ensuremath{\text{W}}\xspace}
\newcommand{\PZ}{\ensuremath{\text{Z}}\xspace}

\newcommand{\MW}{\ensuremath{M_\PW}\xspace}

\newcommand{\MZ}{\ensuremath{M_\PZ}\xspace}

\newcommand{\GZ}{\ensuremath{\Gamma_\PZ}\xspace}

\newcommand{\GW}{\ensuremath{\Gamma_\PW}\xspace}

\newcommand{\GeV}{\ensuremath{\,\text{GeV}}\xspace}
\newcommand{\TeV}{\ensuremath{\,\text{TeV}}\xspace}

\newcommand{\GF}{\ensuremath{G_\mu}}

\newcommand{\ptsub}[1]{\ensuremath{p_{\text{T},#1}}\xspace}

\newcommand{\newc}{\newcommand}
\newc{\bi}{\begin{itemize}}
\newc{\ei}{\end{itemize}}
\newc{\benu}{\begin{enumerate}}
\newc{\eenu}{\end{enumerate}}
\newc{\bc}{\begin{center}}
\newc{\ec}{\end{center}}
\newc{\bfig}{\begin{figure}}
\newc{\efig}{\end{figure}}
\newc{\qbar}{\bar{q}}
\newc{\go}{\tilde{g}}
\newc{\PB}{\textsc{Powheg-Box}}

\newcommand{\MoCaNLO}{{\sc MoCaNLO}\xspace}

\newcommand{\madgraphbis}{{\sc\small MG5\_aMC@NLO}\xspace}
\newcommand{\rT}{{\mathrm{T}}}
\newcolumntype{.}{D{.}{.}{-1}}
\newcolumntype{d}[1]{D{.}{.}{#1}}

\newcommand{\EW}{\ensuremath{\text{EW}}}

\newcommand{\mr}[1]{\ensuremath{\mathrm{#1}}}
\newcommand{\mc}[1]{\ensuremath{\mathcal{#1}}}
\newcommand{\muR}{\ensuremath{\mu_{\mr{R}}}}
\newcommand{\muF}{\ensuremath{\mu_{\mr{F}}}}

\colorlet{tableoverheadcolor}{gray!37.5}
\colorlet{tableheadcolor}{gray!25}
\colorlet{tablerowcolor}{gray!12.5}

\newlength{\width}
\newlength{\height}

\def\draftdate{\relax}
\def\mda{\relax}
\def\mua{\relax}
\def\mla{\relax}
\def\draft{
\def\thtystars{******************************}
\def\sixtystars{\thtystars\thtystars}
\typeout{}
\typeout{\sixtystars**}
\typeout{* Draft mode!
         For final version remove \protect\draft\space in source file *}
\typeout{\sixtystars**}
\typeout{}
\def\draftdate{\today}
\def\mua{\marginpar[\boldmath\hfil$\uparrow$]%
                   {\boldmath$\uparrow$\hfil}\color{black}%
                    \typeout{marginpar: $\uparrow$}\ignorespaces}
\def\mda{\color{red}\marginpar[\boldmath\hfil$\downarrow$]%
                   {\boldmath$\downarrow$\hfil}%
                    \typeout{marginpar: $\downarrow$}\ignorespaces}
\def\mla{\marginpar[\boldmath\hfil$\rightarrow$]%
                   {\boldmath$\leftarrow $\hfil}%
                    \typeout{marginpar: $\leftrightarrow$}\ignorespaces}
\def\Mua{\marginpar[\boldmath\hfil$\Uparrow$]%
                   {\boldmath$\Uparrow$\hfil}\color{black}%
                    \typeout{marginpar: $\uparrow$}\ignorespaces}
\def\Mda{\color{red}\marginpar[\boldmath\hfil$\Downarrow$]%
                   {\boldmath$\Downarrow$\hfil}%
                    \typeout{marginpar: $\downarrow$}\ignorespaces}
\def\Mla{\marginpar[\boldmath\hfil\textcolor{red}{$\Rightarrow$}]%
                   {\boldmath\textcolor{red}{$\Leftarrow $}\hfil}%
                    \typeout{marginpar: $\leftrightarrow$}\ignorespaces}
\overfullrule 5pt
\oddsidemargin 15mm
\marginparwidth 29mm
}

\begin{document}

\title{\hfill ~\\[-30mm]
\phantom{h} \hfill\mbox{\small FR-PHENO-2023-05, MPP-2023-122}
\\[1cm]
\vspace{13mm}   \textbf{Amplitude-assisted tagging of longitudinally polarised bosons using wide neural networks}}

\date{}
\author{
Michele Grossi$^{1\,}$\footnote{E-mail:
  \texttt{michele.grossi@cern.ch}},
Massimiliano Incudini$^{2\,}$\footnote{E-mail:
  \texttt{massimiliano.incudini@univr.it}},
Mathieu Pellen$^{3\,}$\footnote{E-mail:
  \texttt{mathieu.pellen@physik.uni-freiburg.de}},
Giovanni Pelliccioli$^{4\,}$\footnote{E-mail: \texttt{gpellicc@mpp.mpg.de} (corresponding author)}
\\[9mm]
{\small\it $^1$ European Organisation for Nuclear Research (CERN)}\\
{\small\it Geneva 1211, Switzerland}\\[3mm]
{\small\it $^2$ Department of Computer Science, University of Verona,} \\
{\small\it Strada Le Grazie 15, 37129 Verona, Italy}\\[3mm]
{\small\it $^3$ Physikalisches Institut, Universit\"at Freiburg,} \\
{\small\it Hermann-Herder-Str. 3, 79104 Freiburg, Germany}\\[3mm]
{\small\it $^4$ Max-Planck-Institut f\"ur Physik,} \\
{\small\it F\"ohringer Ring 6, 80805 M\"unchen, Germany}\\[3mm]
}
\maketitle

\begin{abstract}
\noindent
Extracting longitudinal modes of weak bosons in LHC processes is essential to understand the electroweak-symmetry-breaking mechanism.
To that end, we propose a general method, based on wide neural networks, to properly model longitudinal-boson signals and hence enable the event-by-event tagging of longitudinal bosons.
It combines experimentally accessible kinematic information and genuine theoretical inputs provided by amplitudes in perturbation theory.
As an application we consider the production of a Z boson in association with a jet at the LHC, both at leading order and in the presence of parton-shower effects. The devised neural networks are able to extract reliably the longitudinal contribution to the unpolarised process. 
The proposed method is very general and can be systematically extended to other processes and problems.
\end{abstract}
\thispagestyle{empty}
\vfill
\newpage
\setcounter{page}{1}

\tableofcontents
\newpage

\section{Introduction}\label{sec:intro}
Accessing the polarisation of electroweak (EW) bosons at high-energy colliders is crucial to gain insights in the electroweak-symmetry-breaking mechanism (EWSB), whose nature is currently explained by the Higgs mechanism \cite{Englert:1964et,Higgs:1964pj,Guralnik:1964eu}.
By means of the EWSB, the $\PW$ and $\PZ$ bosons are given a mass and a longitudinal-polarisation state. Therefore, any deviation in the production of longitudinal bosons in scattering processes would suggest the presence of new-physics effects, implying a different realisation of the EWSB compared to the Standard Model (SM) one.

The investigation of polarised-boson signals in LHC processes is becoming an important part of the analysis programme of the ATLAS and CMS collaborations with Run-2 data, as shown in recent measurements of di-boson production and vector-boson scattering (VBS) \cite{Aaboud:2019gxl,Sirunyan:2020gvn,CMS:2021icx,ATLAS:2022oge}.
The increase in statistics of Run-3 and the High-Luminosity phase will drastically improve the precision of current analyses and give access to polarised signals in complex multi-boson processes \cite{CMS:2018mbt,Azzi:2019yne,Roloff:2021kdu}.

The analysis paradigm for the measurement of polarisations of EW bosons with Run-1 data at $7/8\TeV$ was the evaluation of angular coefficients of the boson decay rate, which are related to the polarisation fractions. 
The extraction of angular coefficients in LHC processes was proposed in seminal phenomenological works \cite{Bern:2011ie,Stirling:2012zt,Belyaev:2013nla} 
and applied in experimental analyses of 
$\PW+$j \cite{Chatrchyan:2011ig,ATLAS:2012au} and
$\PZ+$j \cite{Khachatryan:2015paa,Aad:2016izn} events, as well as of top-quark decays \cite{ATLAS:2016fbc,Khachatryan:2016fky,CMS:2020ezf}.
Owing to its simplicity, this strategy has been further investigated and extended in more recent phenomenological studies \cite{Gauld:2017tww,Baglio:2018rcu,Rahaman:2018ujg,Rahaman:2019lab,Baglio:2019nmc,Frederix:2020nyw,Rahaman:2021fcz,Pellen:2022fom}. However, its application is limited \cite{Bern:2011ie,Stirling:2012zt,Belyaev:2013nla,Baglio:2019nmc,Frederix:2020nyw,Ballestrero:2017bxn}
to inclusive decays (\ie without selections on single decay products of bosons) and to two-body decays [\ie without radiative corrections to the decay (\emph{e.g.}\ EW corrections)]. 

More recently, a different approach was proposed \cite{Ballestrero:2017bxn} to interpret LHC Run-2 data in terms of polarisations of EW bosons, relying on the direct simulation of intermediate polarised bosons in Monte Carlo codes. 
This method has been automated and applied to several processes
at leading order \cite{Ballestrero:2017bxn,Ballestrero:2019qoy,Ballestrero:2020qgv,BuarqueFranzosi:2019boy}
and extended to higher-order EW and QCD corrections \cite{Denner:2020bcz,Denner:2020eck,Poncelet:2021jmj,Denner:2021csi,Pellen:2021vpi,Le:2022lrp,Le:2022ppa}.
The existing results in the literature in this direction concern 
vector-boson scattering (VBS) \cite{Ballestrero:2017bxn,Ballestrero:2019qoy,Ballestrero:2020qgv,BuarqueFranzosi:2019boy}, 
inclusive di-boson \cite{Denner:2020bcz,Denner:2020eck,Poncelet:2021jmj,Denner:2021csi,Le:2022lrp,Le:2022ppa,Denner:2022riz}, 
Higgs-boson decays \cite{Brehmer:2014pka,Maina:2020rgd,Maina:2021xpe} 
and $\PW+$j production \cite{Pellen:2021vpi}. 
The great advantage of this approach is that, upon subtraction of backgrounds, LHC signal events can be fitted with 
polarised templates in order to extract polarisation fractions in a differential 
way from the LHC data, accounting properly for interference effects and spin correlations. This has become the new
analysis paradigm with Run-2 data, as demonstrated by the pioneering measurements performed by ATLAS and CMS in di-boson inclusive production \cite{Aaboud:2019gxl,CMS:2021icx,ATLAS:2022oge} and VBS \cite{Sirunyan:2020gvn}.

A number of recent studies have been carried out targeting the 
polarisation extraction in the presence of hadronic decays of
the weak bosons \cite{De:2020iwq,Grossi:2020orx,Kim:2021gtv,Dey:2021sug,Ricci:2022htc,Denner:2022riz}, both with the polarised-template method and with machine-learning techniques.
The usage of machine learning was also proposed to extract polarisation fractions in VBS, starting from the kinematic structure of the events \cite{Searcy:2015apa,Lee:2018xtt,Lee:2019nhm,Li:2021cbp,Ozdemir:2023rya,Grossi:2020orx}.

Independently of the specific approach that may be used for the interpretation, the polarisation state of an unstable particle, like EW bosons, is not directly accessible in the detectors. Therefore, the information about it can only be reconstructed (in a probabilistic way) from the stable decay products. In other words, the polarisation of EW bosons is a pseudo-observable.
%%%%%
On the other hand, from a theoretical perspective, the whole information regarding the fundamental quantum field theory and therefore the polarisation properties is encoded in the amplitude. 
Therefore, accessing the amplitude of scattering processes at experiments would give the maximal information possible, \ie the maximal predictive power. However, in a realistic experimental environment only the momenta of the visible final states can be reconstructed. This means that the momenta of the initial states as well as their parton type, both needed for the exact evaluation of the amplitude, is a priori unknown.
As explained in the rest of the article, machine learning (ML) can actually be used to approximate well the amplitudes based only on the partial information available experimentally.

It is worth pointing out that in this article we exclusively refer to \emph{amplitudes} and not \emph{matrix elements}, which are actually equivalent for our purposes. Our method, though close in spirit, should not to be confused with the \emph{matrix-element method} \cite{Kondo:1988yd,Kondo:1991dw} and the 
\emph{optimal-observable method} \cite{Diehl:1993br,Diehl:1996wm,Janot:2015yza}.

The article is organised as follows. 
In Sect.~\ref{sec:what_about} we explain the difficulties in tagging longitudinal bosons and propose a solution based on wide neural networks and amplitudes. A concrete application of the proposed method to $\PZ+\Pj$ at the LHC is then detailed and discussed in Sect.~\ref{sec:applic}. In Sect.~\ref{sec:concl} we draw the conclusions of our work.

\section{Tagging longitudinal bosons}\label{sec:what_about}
\subsection{Definition of the problem}\label{sec:def_of_probl}
A generic (unpolarised) amplitude featuring a resonant gauge boson decaying into a lepton-neutrino pair can be written as follows (in the unitary gauge),  % \MP{make formula general}
\begin{equation}\label{eq:Mlep}
\mathcal{M} = \mathcal{M}^{\mc P}_{\mu} \frac{i}{k^2 - M_V^2 + i \Gamma_V M_V}\left(-g^{\mu\nu}+\frac{k^{\mu}
  k^{\nu}}{M_V^2}\right)
  \mathcal{M}^{\mc D}_{\nu} ,
%\left\{\frac{-i\,g}{\sqrt{2}}\bar{\psi}_{\ell} \gamma_{\nu}
%\bigg[
%P_{\rm L}\left(\frac{1 - \gamma^5}2\right)+
%P_{\rm R}\left(\frac{1 + \gamma^5}2\right)
%\bigg]
%\psi_{\nu_{\ell}}
%\right\} ,
\end{equation}
where $\mc{M}^{\mc P}$ and $\mc{M}^{\mc D}$ describe the production and decay part of the amplitude, respectively. The quantities
$M_V$ and $\Gamma_V$ represent the gauge-boson mass and width, respectively.
In particular, the tensor part of the propagator can be cast into the following form,
\begin{equation}
-g^{\mu\nu} + \frac{k^{\mu}k^{\nu}}{M^2} = \sum_{\lambda = 1}^4 \varepsilon^{\mu}_\lambda(k)
\varepsilon^{\nu^*}_{\lambda}(k)\,\,,
\end{equation}
where the $\{\varepsilon^{\mu}_\lambda(k)\}$ represent polarisation vectors of the massive gauge boson.
The sum runs over four polarisation states, namely the three physical states and a fourth one, whose
structure depends on the EW-gauge choice, and is thus is unphysical\footnote{It is important to notice that this unphysical contribution always cancels out against Goldstone-boson contributions at any order in perturbation theory.}.
Throughout the article, we use the labels $\rL$, $+$, and $-$ for the longitudinal, right-handed, and left-handed
states, respectively.
Notice that the polarisation vectors are defined in such a way that they are transverse w.r.t.\ the boson four momentum, but do not transform as Lorentz covariants, therefore they must be defined in a
specific Lorentz frame.
%\MP{@Giovanni: Do we specify anywhere what frame we use?}
Also, we would like to emphasise that the definition of the polarisation of the massive gauge bosons is only meaningful when gauge bosons are on the mass shell or when the resonant contributions are treated in the narrow-width \cite{Uhlemann:2008pm,Artoisenet:2012st} or pole approximation \cite{Stuart:1991cc,Stuart:1991xk,Aeppli:1993rs,Aeppli:1993cb,Denner:2000bj,Denner:2005fg}. The reason for this is to guarantee gauge invariance.

The amplitude in Eq.~\eqref{eq:Mlep}, including both production and decay parts, can therefore be written as,
\begin{equation}\label{eq:Msum}
\mathcal{M} = \sum_{\lambda} \mathcal{M}_\lambda\,,\qquad \lambda=\rL, +,- \, ,
\end{equation}
where $\mathcal{M}_\lambda$ is the amplitude with a polarised intermediate gauge boson (with state $\lambda$),
\begin{equation}\label{eq:Mlep_pol_lambda}
\mathcal{M}_\lambda = \big[\mathcal{M}^{\mc P}_{\mu} \varepsilon^{\mu}_\lambda(k)\big] \frac{i}{k^2 - M_V^2 + i \Gamma_V M_V}
\big[
\varepsilon^{\nu^*}_{\lambda}(k)
  \mathcal{M}^{\mc D}_{\nu} \big]\,.
\end{equation}
Hence, squaring the unpolarised amplitude leads to
\begin{equation}
\label{eq:interfpol}
\left|\mathcal{M}\right|^2 = {\sum_{\lambda}\left|
  \mathcal{M}_{\lambda}\right|^2} + {\sum_{\lambda \neq \lambda'}
  \mathcal{M}_{\lambda}^{ *}\mathcal{M}_{\lambda'}}\,,\qquad \lambda,\lambda'=\rL, +,-\,,
\end{equation}
where the first sum represents the incoherent sum over polarised squared amplitudes,
while the second one includes all interference terms.
For phenomenological purposes it is convenient \cite{Ballestrero:2019qoy} to define a transverse ($\rT$) contribution as
the coherent sum of the left- and right-handed contributions to the squared amplitude,
\begin{equation}
  \left|\mathcal{M}_{\rT}\right|^2 =
  \left|\mathcal{M}_{+}\right|^2 + \left|\mathcal{M}_{-}\right|^2
  +
  2\,{\rm Re}\,\left(\mathcal{M}_{+}^{ *}\mathcal{M}_{-}\right)\,,
\end{equation}
leading to a simpler structure of Eq.~\eqref{eq:interfpol},
\begin{equation}\
  \label{eq:L+T}
  \left|\mathcal{M}\right|^2 =
  \left|\mathcal{M}_{\rL}\right|^2 + \left|\mathcal{M}_{\rT}\right|^2
  +
  2\,{\rm Re}\,\left(\mathcal{M}_{\rL}^{ *}\mathcal{M}_{+}\right)
  +
  2\,{\rm Re}\,\left(\mathcal{M}_{\rL}^{ *}\mathcal{M}_{-}\right)\,.
\end{equation}
The term $\left|\mathcal{M}_{\rL}\right|^2$ defines the longitudinally  polarised squared amplitude which is the focus of the present work.
Note that the interference terms of Eq.~\eqref{eq:interfpol} [or of Eq.~\eqref{eq:L+T}] are in general non vanishing and can take either positive or negative values.

Hence, the fully differential unpolarised and longitudinal cross sections schematically read,
\begin{align}
{\rm d} \sigma_{\rm unp} = \frac{1}{F} \,\left|\mathcal{M}\right|^2 \,{\rm d} \Phi,\qquad 
{\rm d} \sigma_{\rL} = \frac{1}{F} \, \left|\mathcal{M}_{\rL}\right|^2 \, {\rm d} \Phi\,,
\end{align}
where the flux factor and phase-space measure are denoted by $F$ and ${\rm d} \Phi$, respectively. The differential longitudinal fraction in a generic observable $\mathcal{O}$, that is ought to be extracted experimentally, is therefore defined as,
\begin{align}
\label{eq:fraction}
 f_{\rL}(\mathcal{O})= \frac{{\rm d} \sigma_{\rL}}{\rm d \mathcal{O}}\bigg/\frac{\rm d \sigma_{\rm unp}}{\rm d \mathcal{O}} \,.
\end{align}
%
%All these definitions hold both at the differential level and for the total cross sections.
\paragraph{Problem}
The main challenge is to extract the longitudinal fraction experimentally for arbitrary observables, \ie in a fully differential way.
In other words, we would like to answer the question: how can we infer on an event-by-event basis the probability for an LHC event to be longitudinally polarised? In what follows, we will address exactly this problem.
\\

As briefly mentioned in Sect.~\ref{sec:intro}, a number of methods have been proposed in past and recent years to address the issue of longitudinal-event tagging.
In the rest of this section, we review some of them.

For the interpretation of LHC Run-1 data, the \emph{angular-coefficient method} was typically applied with polarisation-extraction purposes. It relies on the functional structure of the tree-level decay rate of EW bosons, that can be written as follows \cite{Bern:2011ie,Stirling:2012zt},
\begin{eqnarray}
\label{eq:Vdecll}
\frac{\rd^3\sigma}{\rd\cos\theta^*\,\rd\phi^*\,\rd \mc O} 
&=&
\frac{\rd \sigma}{\rd \mc O}\frac{3}{16\pi}
\bigg[
1+\cos^2\theta^* + A_0\frac{1-3\cos^2\theta^*}2 + A_4\cos\theta^* 
+ 2A_1 \cos\theta^*\sin\theta^*\cos\phi^*\nnb\\
&&\hspace{1.4cm}+ \; 2A_6 \cos\theta^*\sin\theta^*\sin\phi^* 
+ A_3 \sin\theta^*\cos\phi^*
+ A_5 \sin\theta^*\sin\phi^*\nnb\\
&&\hspace{1.4cm}+\;\frac12 A_2 \sin^2\theta^*\cos2\phi^* +  A_7 \sin^2\theta^*\sin2\phi^*\bigg]\,,
\end{eqnarray}
where $\theta^*$ and $\phi^*$ are the polar and azimuthal decay angle of a decay lepton in the decayed-boson rest frame, calculated w.r.t.\ the boson trajectory in a certain Lorentz frame (the one where polarisation states are defined).
The coefficients $\{A_0,\ldots,A_7\}$, which are functions of the observable $\mathcal{O}$ (independent of decay angles), are related to polarisation fractions $f_\rL$ and $f_\pm$ via linear combinations. 
Projecting Eq.~\eqref{eq:Vdecll} onto suitable spherical harmonics of rank 2, the polarisation fractions can be then easily extracted. This strategy is valid for a single boson, in the absence of radiative corrections to the decay, and in a fully inclusive decay-phase-space measure, \ie without any cut on individual decay products. Applying the projections in the presence of transverse-momentum and/or rapidity cuts on decay products, as done in many experimental analyses, may give results that are far from describing the polarisation structure of a process \cite{Stirling:2012zt,Belyaev:2013nla,Ballestrero:2017bxn,Baglio:2018rcu,Frederix:2020nyw}. Extensions of the method to account for multi-boson spin correlations \cite{Rahaman:2018ujg,Rahaman:2019lab,Rahaman:2021fcz,Boudjema:2009fz} are also limited due to similar reasons.

The extraction of angular coefficients from decay rates can also be applied differentially in any LHC observables, providing a way to reweight unpolarised LHC events according to polarisation fractions and therefore split the events into longitudinal and transverse samples. This approximate method has been applied for Run-1 $V+\Pj$ events \cite{Chatrchyan:2011ig,ATLAS:2012au,Khachatryan:2015paa,Aad:2016izn} 
but was proven to fail in certain kinematic regimes, especially due to the wrong assumption that polarisation fractions are the same in the presence and absence of decay-product selections \cite{Ballestrero:2019qoy}.

The most prominent way to extract polarised signals out of Run-2 LHC data is the so-called \emph{polarised-template method}.
Building on a theoretically sound definition of polarised signals at
amplitude level \cite{Ballestrero:2017bxn} that can be systematically extended to higher orders in perturbation theory \cite{Denner:2020bcz,Denner:2020eck,Poncelet:2021jmj,Denner:2021csi,Pellen:2021vpi,Le:2022lrp,Le:2022ppa}, the method relies on separate templates for each physical polarisation state and for the interference terms. Upon a previous subtraction of reducible and irreducible backgrounds, the (unpolarised) signal events are simultaneously fitted with fully independent polarised templates. This can be applied differentially to any LHC observable. 
In practice, these fits are restricted to those observables that are thought to be the most sensitive to discriminate between longitudinal and transverse modes. This does not guarantee that the polarisation information is fully exploited from the available data.
    In addition, the fitting procedure requires quite intensive theoretical calculations that only recently achieved (N)NLO accuracy \cite{Denner:2020bcz,Denner:2020eck,Poncelet:2021jmj,Denner:2021csi,Pellen:2021vpi,Le:2022lrp,Le:2022ppa}. To be of any use, such calculations should be performed in a fiducial phase-space volume which is exactly the one used in the experimental analysis, a task that can turn out to be far from trivial.

The idea of using ML methods to facilitate the extraction of polarisation fractions has been already explored in the literature. It has been applied in the case of EW bosons produced in VBS and inclusive di-boson production, both with leptonic \cite{Searcy:2015apa,Lee:2018xtt,Lee:2019nhm,Grossi:2020orx,Li:2021cbp,Ozdemir:2023rya} and hadronic decays \cite{Grossi:2020orx,Kim:2021gtv}.  The proposed ML approaches typically rely on kinematic observables approximating decay angles in the case of leptonic decays of $\PW$ bosons, and on jet-substructure observables to treat hadronic decays, with the aim of performing an event-by-event classification, possibly accounting for new-physics effects that may distort the underlying dynamics.

The method we propose to tag longitudinal bosons lies somehow at the intersection amongst the aforementioned methods, complementing accessible kinematic information of LHC events with a genuine theoretical input given by amplitudes describing the process dynamics.

As a last remark before detailing our strategy, we stress that the polarisation structure of a process is model dependent, owing to possibly different dynamics at production and decay level. 
The advantage of the solution we present in the following is that the model dependence is uniquely encoded in the amplitudes.

\subsection{A machine-learning-based solution}
\label{sec:sol_prob}
Equation~\eqref{eq:fraction} implies that at the phase-space--point level, the polarisation fraction is equal to the ratio of the longitudinally polarised squared amplitude over the unpolarised one,
\begin{align}
\label{eq:rL}
 r_\rL = \frac{\left|\mathcal{M}_\rL \right|^2}{\left|\mathcal{M}\right|^2} .
\end{align}
This statement is exact at leading order (LO) in perturbation theory for each partonic channel occurring in the process.
It means that computing $r_\rL$ for a given process requires the knowledge of the full kinematics as well as of the
flavours of the partonic-process external particles.

Considering an unpolarised event sample, this ratio can be computed for each event separately. 
One can therefore tag each event as longitudinally polarised or not by sampling on the value of $r_\rL$.
From the unpolarised event sample, one can therefore obtain a longitudinally polarised sample.
This procedure is completely equivalent to generating a longitudinal sample from scratch.
It follows that the fully differential knowledge of $r_\rL$ allows for the assessment of longitudinal polarisation on a even-by-event basis.
We note that while Eq.~\eqref{eq:interfpol} does not guarantee this ratio to be comprised between $0$ and $1$, in practice it is and allows a straightforward sampling without requiring to determine the minimum and maximum of $r_\rL$ beforehand.

As mentioned above, this procedure is exact at LO accuracy.
It can actually be extended to a LO sample with parton-shower (PS) corrections using a similar procedure. Considering an unpolarised sample at LO+PS accuracy, one can compute $r_\rL$ with the original event (before showering) and tag the event after showering based on the value of $r_\rL$.
Again, this procedure is equivalent to generate a longitudinal sample at LO+PS accuracy from scratch\footnote{Note that instead of Eq.~\eqref{eq:rL}, we have also considered a variation of it, namely: $r_{{\rm D},\rL} = \left|\mathcal{M}_{{\rm D},\rL}\right|^2/\left|\mathcal{M}_{\rm D}\right|^2$, where the subscript ${\rm D}$ denotes the decay of the gauge bosons.
It turned out that this variable is not sufficient to reproduce the full longitudinal fraction.}.

\paragraph{Key concept}
Given the possibility to evaluate $r_\rL$, one can tag events as longitudinally polarised.
While doable theoretically, this is unfortunately not possible experimentally as the evaluation of $r_\rL$ at the event level requires the knowledge of all momenta and flavours of the initial and final partons, an information which is not available experimentally.
Instead, what is available experimentally is the knowledge of the final-state momenta which constitutes therefore only a partial information.
The central idea is therefore to bypass this lack of information by using a neural network (NN) to obtain an approximate value of $r_\rL$ called $\tilde r_\rL$ which in turn depends only on the experimental information available.
In other words, the \NN is trained to mimic $r_\rL$ with an incomplete information.
Later, we show that this method is applicable in practice.
\\

It implies, therefore, that one can use $\tilde r_\rL$ to tag experimental events as longitudinal.
The longitudinal fraction extracted in this way can then be compared against theoretical predictions.
If $r_\rL$ is computed within the SM, an agreement between the theoretical predictions and the extracted value of the fraction indicates that the data is compatible with SM expectations.
On the other hand, a disagreement would be the sign of a failure of the SM to describe the physics at hand.
The procedure can be applied not only to the SM but also to any \emph{UV-complete model} as well as to rather model-independent frameworks like \emph{simplified models} or \emph{effective-field theories}.

The method we propose does not require any fitting procedure.
It is by definition multi-dimensional and therefore ensures that all possible information available experimentally is used.
It is also very flexible with respect to the phase-space requirements. In fact, if the training is done in an inclusive phase space, the trained model can be used in any fiducial volume that is more restrictive.

In summary, the key idea of our approach is to relate the tagging of LHC events to a single theoretically clean quantity, using machine learning to cope with incomplete information.

\section{Application: Z+jet}\label{sec:applic}
In order to illustrate the newly devised method, we apply it to the extraction of the longitudinal polarisation of a $\PZ$ boson in $\PZ+\Pj$ production at the LHC.
We would like to emphasis that, in spite of possibly different input features for the training of the \NN for a different process, our method is fully general and can be applied to any process featuring one or several $\PZ$ or $\PW$ boson(s).
The process we consider is, 
\begin{equation}\label{eq:resproc}
\Pp\Pp\to\Pj+\PZ(\to\mu^+\mu^-)\,+X\,.
\end{equation}
While providing a non-trivial test-bed, this reaction is particularly suited for polarisation studies as it has a very high cross section and allows for the full reconstruction of the final state.

Note that the production of a muon-antimuon pair is mediated by a photon and a $\PZ$ boson. However, since we aim at extracting the polarisation of an intermediate $\PZ$ boson, the photon contribution is regarded as an irreducible background to be subtracted before any polarisation analysis \cite{Denner:2021csi,Denner:2022riz}. 
In the presence of a cut on the lepton-pair invariant mass around the $\PZ$ pole mass, the photon background (as well as the photon-$\PZ$ interference) is typically small.
In the setups considered here (see Sect.~\ref{sec:setup}), this irreducible background is at the level of $1\%$, estimated from a comparison between the $\PZ$-mediated signal of Eq.~\eqref{eq:resproc} and the full off-shell calculation of $\Pp\Pp\rightarrow \Pj+\mu^+\mu^-$.

Since we are interested in polarised signals, we choose to define polarisation vectors in the Lorentz frame where the $\PZ$ boson and the jet are back to back, which coincides (at LO) with the partonic centre-of-mass frame. This reference frame is the one where the $2\rightarrow 2$ scattering happens and can be entirely reconstructed up to experimental uncertainties.
Therefore, this choice is well motivated both from a theoretical and from an experimental viewpoint.
\change{
  We stress that any polarisation extraction from simulated events or experimental data is frame dependent.
  This means that, although the general strategy we propose can be applied to any polarisation-frame definition,
  the application considered here depends on the specific choice of the polarisation frame.
  In practice, the $r_{\rL}$ quantity defined in Eq.~\eqref{eq:rL} takes different values when computed for the same phase-space point
  but for different polarisation-frame choices, therefore the NN-training stage is tailored to the specific choice of
  polarisation frame.
}

\subsection{Input parameters and event selections}\label{sec:setup}
In this section, we list the input SM parameters used for the numerical computations and the event selections considered for the phenomenological analysis.

The simulations are performed at a centre-of-mass energy of $\sqrt{s} = 13.6 \TeV$ for proton-proton collisions at the LHC.
The parton distribution function \texttt{NNPDF31\_nlo\_as\_0118}~\cite{NNPDF:2017mvq} has been utilised thanks to \textsc{Lhapdf}~\cite{Buckley:2014ana}.
The renormalisation and factorisation scales are fixed to
\begin{equation}
\label{eq:scale}
 \muR = \muF = \MZ .
\end{equation}
The \EW\ coupling is fixed through the $G_\mu$ scheme \cite{Denner:2000bj,Dittmaier:2001ay} is used for the electroweak coupling as
\begin{equation}
  \alpha = \frac{\sqrt{2}}{\pi} G_\mu \MW^2 \left( 1 - \frac{\MW^2}{\MZ^2} \right)  \qquad \text{with}  \qquad   \GF = 1.16639\times 10^{-5}\GeV^{-2}\;.
\end{equation}
The following masses and widths have been taken,
\begin{alignat}{2}
                \MZ &=  91.188\GeV,      & \quad \quad \quad \GZ &= 2.49877\GeV,  \nonumber \\
                \MW &=  80.419\GeV,       & \GW &= 2.09291\GeV .
\end{alignat}
The masses or widths of all other particles do not play a role in this process or have been set to zero. Note that these parameters are essentially the default ones in \madgraphbis~\cite{Alwall:2014hca}.

A number of different event selections are used in this work. 
\noindent
The first one, which we label \emph{generation-level}, is characterised by a transverse-momentum and rapidity cut on the leading jet, as well as an invariant-mass cut on the charged-lepton pair,
\begin{align}
\label{eq:jet_incl}
 \ptsub{\Pj} >  10\GeV\,,\qquad  
 |y_{\Pj}| < 5, \qquad \textrm{and} \qquad 
 76\GeV < M_{\mu^+\mu^-} < 106\GeV\,.
\end{align}
With this setup we have generated the initial parton-level event samples for both unpolarised and longitudinally polarised $\PZ$ bosons. 

The second selection, labeled \emph{inclusive}, used for some of the phenomenological results with and without PS effects, is characterised by slightly more restrictive cuts, in order to avoid biasing the PS application, namely,
\begin{align}
\label{eq:inclusive_setup}
 \ptsub{\Pj} >  20\GeV\,, \qquad |y_{\Pj}| < 4 ,
 \qquad \textrm{and} \qquad 81\GeV < M_{\mu^+\mu^-} < 101\GeV\,.
\end{align}
Notice that both the \emph{generation-level} and \emph{inclusive} setups avoid any additional cut on the $\PZ$-boson decay products, making the selections not realistic in a collider environment. However, it enables to be as inclusive as possible for the training of NNs, ensuring that any realistic selection will be enclosed in the phase-space region.

Finally, the third selection, which we dub \emph{fiducial}, is then used to mimic a realistic setup at the LHC. In addition to the cuts in Eq.~\eqref{eq:inclusive_setup}, transverse-momentum and rapidity cuts are applied on the charged leptons,
\begin{align}
\label{eq:fiducial_setup}
 \ptsub{\mu^\pm} >  20\GeV \qquad \textrm{and} \qquad |y_{\mu^\pm}| < 2.7\,.
\end{align}

\subsection{Tools}
For the generation of longitudinal and unpolarised parton-level events, we have used version 2.7.3 
 of \madgraphbis \cite{BuarqueFranzosi:2019boy}, 
which enables to select intermediate-resonance
helicity states in the narrow-width approximation \cite{Artoisenet:2012st}.
As a validation of the longitudinal signal, we have compared the \madgraphbis results against those obtained with the private Monte Carlo framework \MoCaNLO, that uses the pole-approximation approach detailed in Refs.~\cite{Denner:2016wet,Denner:2020bcz,Denner:2020eck,Denner:2021csi,Denner:2022riz}. Good agreement has been found.

In order to compute PS effects, we have used version 8.244 of the {\sc Pythia8} program \cite{Sjostrand:2014zea} with standard settings.  
The space-like and time-like shower have been applied with both QCD and QED effects. For what concerns QED effects, we veto further photon splittings into fermion pairs in the shower. Note that we have not included multi-parton interactions and hadronisation effects. The reason is to keep this example, while non-trivial, as simple as possible. We argue that including these extra effects could simply be included upon performing a new training of the \NN.
The principle of the ML technique we propose would not be hampered by different scale choices, matching and PS settings that may be needed especially when including higher-order effects.
 
Finally, in order to compute $r_\rL$ for the various partonic channels, we have used the matrix-element provider {\sc Recola 1}~\cite{Actis:2012qn,Actis:2016mpe}
\footnote{
We used the most recent {\sc Recola 1} release (version 1.4.3) which supports helicity selections for intermediate resonances at tree and one-loop level. Further documentation on the usage of polarisation-related subroutines can be found at \url{https://recola.gitlab.io/recola2/api/polsel.html\#polsel}.}.

\begin{figure}[htbp]
    \centering
    \includegraphics{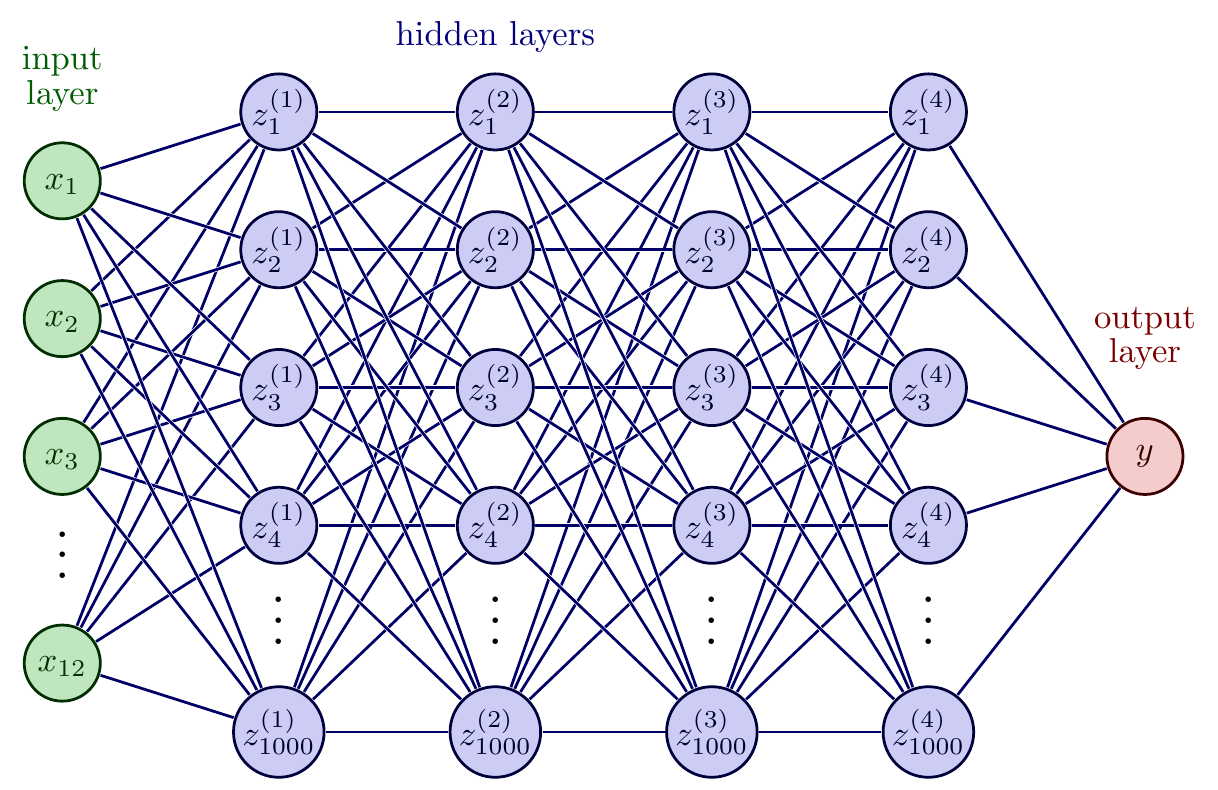}
    \caption{Structure of the neural network. Picture adapted from \citere{nnscheme}.}
    \label{fig:neural-network}
\end{figure}

To obtain an approximate value $\tilde r_\rL$, a machine learning approach was employed, utilising a feed-forward \NN. 
The model is built from the dataset characterised by the following twelve input features, namely the four momenta of the (leading) jet, the antimuon, and the muon:
\begin{equation}
    E_{\Pj},\; p_{x,\Pj},\; p_{y,\Pj},\; p_{z,\Pj},\;
    E_{\mu^+},\; p_{x,\mu^+},\; p_{y,\mu^+},\; p_{z,\mu^+},\;
    E_{\mu^-},\; p_{x,\mu^-},\; p_{y,\mu^-},\; p_{z,\mu^-}\,,
\end{equation}
along with the quantity $r_{\rL}$ defined in Eq.~\eqref{eq:rL}, representing the continuous label. 
%'Emu+', 'pxmu+', 'pymu+', 'pzmu+', 'Emu-', 'pxmu-', 'pymu-', 'pzmu-', 'Ej', 'pxj', 'pyj', and 'pzj', 

The dataset used for 
training and testing containes $286\,073$ and $285\,187$ elements, respectively. As part of the data-preparation process, the dataset has been standardised according to a general procedure where each feature 
is normalised, with the subtraction of the average value of the feature and divided by its standard deviation. 
This approach is more suited than a min-max normalisation, due to its lower sensitivity to outliers.
%This approach is more suited in this case with respect to min-max normalisation, due to its lower sensitivity to outliers.
\noindent
The architecture of the \NN is wide, and this represents a crucial aspect of the proposed technique. Four hidden layers, consisting of 1000 nodes each, were employed. The mathematical 
representation of this \NN involves a series of transformations. The input $x$ has a 
dimensionality of $\R^{12\times 1}$. The subsequent layers, indexed as $i$ ranging from 0 to 4, 
were computed using the formula,
\begin{align}
    z^{(0)}(x; W) & = x , \nonumber \\
    z^{(i+1)}(x; W) & = W_{i+1} \sigma(z^{(i)}(x; W)) + b_{i+1}, \nonumber \\
    y(x; W) & = z^{(5)}(x; W).
\end{align}
The weights are represented by $W = (W_1, W_2, W_3, W_4, W_5, b_1, b_2, b_3, b_4, b_5)$, where 
$W_1 \in \R^{1000 \times 12}$, while 
$W_2, W_3, W_4 \in \R^{1000 \times 1000}$, and $W_5 \in \R^{1 \times 1000}$. 
Additionally, 
$b_1, b_2, b_3, b_4 \in \R^{1000 \times 1}$,
and $b_5 \in \R^{1 \times 1000}$. 
The activation function used was the Rectified Linear Unit (ReLU) \cite{glorot2011deep}, defined as $\sigma(x) = \max\{0, x\}$. The structure of this \NN is depicted in Figure~\ref{fig:neural-network}. 

The design of the machine learning model in this study follows the principles outlined in Ref.~\cite{roberts2022principles}. According to the theory presented, the effectiveness of a \NN is influenced by the dynamics of its training process. It suggests that NNs with a ``deep and narrow'' architecture exhibit chaotic dynamics during training, while those with a ``shallow and wide'' architecture are easier to train. In the asymptotic case, infinitely wide NNs possess a convex loss landscape, enabling the optimal solution to be found through gradient descent. However, such models essentially become linear, losing the non-linear expressivity of the original network and potentially limiting its representational capacity. Therefore, a compromise must be made between the ease of training and the network's expressivity. In our experiments, we achieved satisfactory performance by employing a wide \NN with a width-to-depth ratio of $200$. Such a choice has been made by empirical experimentation and refinement.

The training of the \NN model was performed using the RMSprop algorithm \cite{ruder2016overview}, an adaptive learning rate 
optimisation method specifically designed for mini-batch learning. The algorithm's parameters were set as 
follows: learning rate $\eta = 0.001$, smoothing constant $\alpha = 0.99$, weight decay of $0$, and momentum of $0$. 
The training process spanned $1000$ epochs, with batches of size $500$. 
The code implementing the model was developed in Python 3, making use of the PyTorch library \cite{Paszke_PyTorch_An_Imperative_2019}.

\subsection{Results}

\begin{figure}
\centering
\includegraphics[width=0.6\textwidth]{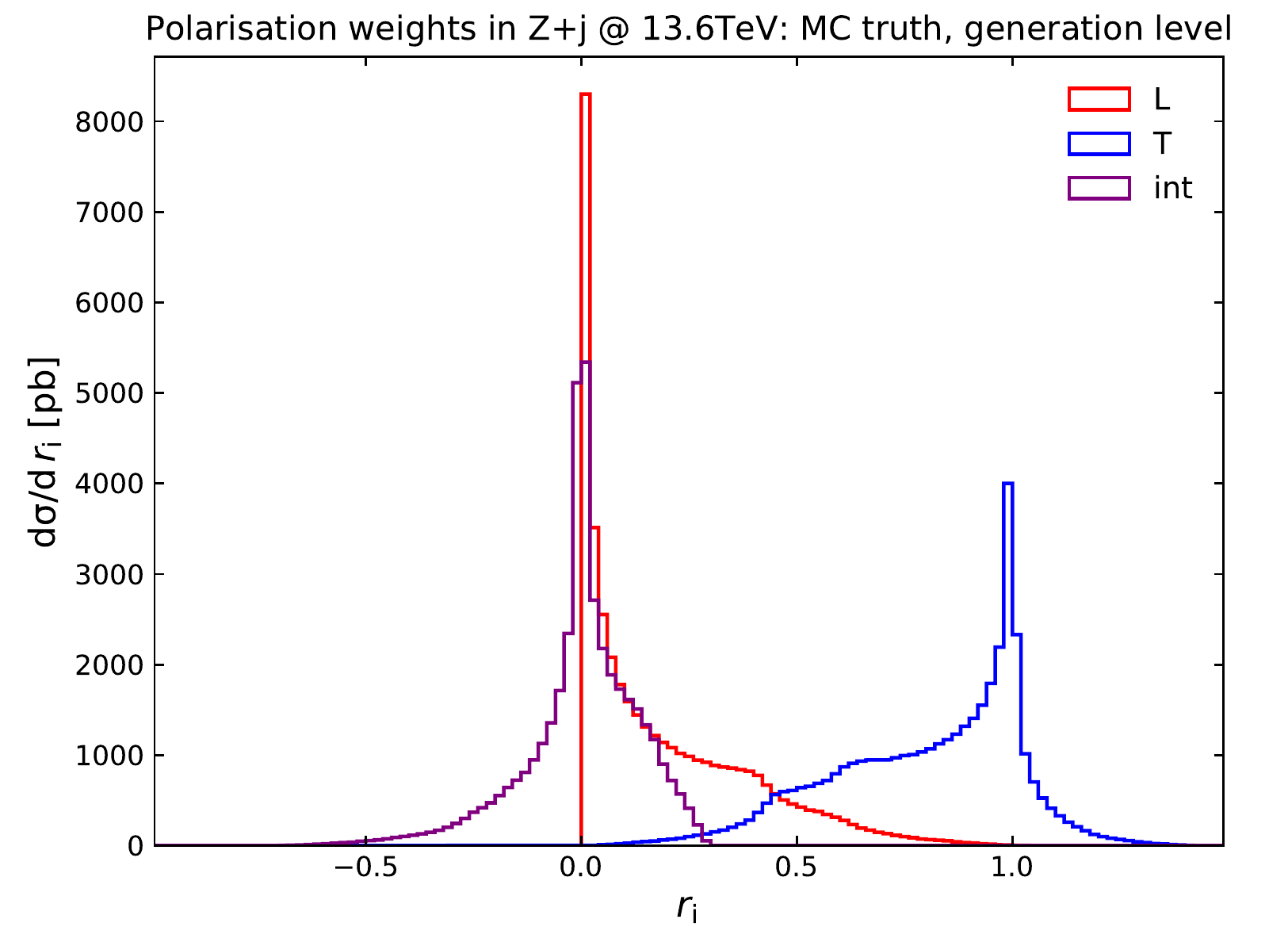}
\caption{\label{fig:rl}
Distributions in the $r_\rL$, $r_\rT$ and $r_{\rm int}$ quantities defined in Eqs.~\eqref{eq:rL} and \eqref{eq:rt_rint}, all normalised to the unpolarised total cross section. The generation-level setup is understood.}
\end{figure}

The key quantity in this application is the ratio $r_\rL$ defined in Eq.~\eqref{eq:rL}. 
It encodes the whole information on the longitudinal-polarisation dynamics, relatively to the polarisation balance in the unpolarised process.
As such, it determines the shape and normalisation of the kinematic distributions for the longitudinal signal.
It is therefore a multi-dimensional function with as many dimensions as the number of random variables needed to generate the momenta of the full final state.
As explained above, this ratio is actually different for each partonic channel.
It means that in order to evaluate it on an event-by-event basis, one does not only need the full kinematic but also the knowledge of the partonic channel.
In order to have a feeling about the structure of $r_\rL$, we show in Fig.~\ref{fig:rl} the differential distribution in $r_\rL$ for unpolarised events, in the generation-level setup. 
We also show, for comparison purposes, the distribution in the corresponding $r_\rT$ and $r_{\rm int}$ ratios, that are respectively defined as,
\begin{equation}\label{eq:rt_rint}
    r_\rT = \frac{|\mc M_{\rT}|^2}{|\mc M|^2}\quad \textrm{and}\quad 
    r_{\rm int} = \frac{2\,{\rm Re}\,(\mc M_{\rL}^*\mc M_{\rT})}{|\mc M|^2} = 1-r_\rL-r_\rT\,.
\end{equation}
From Eq.~\eqref{eq:rt_rint} and Fig.~\ref{fig:rl}, it is clear that, since all amplitudes are complex numbers, the interference term can take negative values, and both $r_\rL$ and $r_\rT$ can exceed the unit. Owing to a peak at 1, the $r_\rT$ distribution in Fig.~\ref{fig:rl} suggests that in the considered process the transverse-polarisation component is way larger than the longitudinal one. This is a well-known result in the SM \cite{Bern:2011ie, Stirling:2012zt}.
It also shows that the proposed method is particularly efficient as it can make full use of this discriminating power.

\begin{figure}
        \setlength{\parskip}{-10pt}
        \begin{subfigure}{0.49\textwidth}
%                 \subcaption{}
                 \includegraphics[width=\textwidth,page=9]{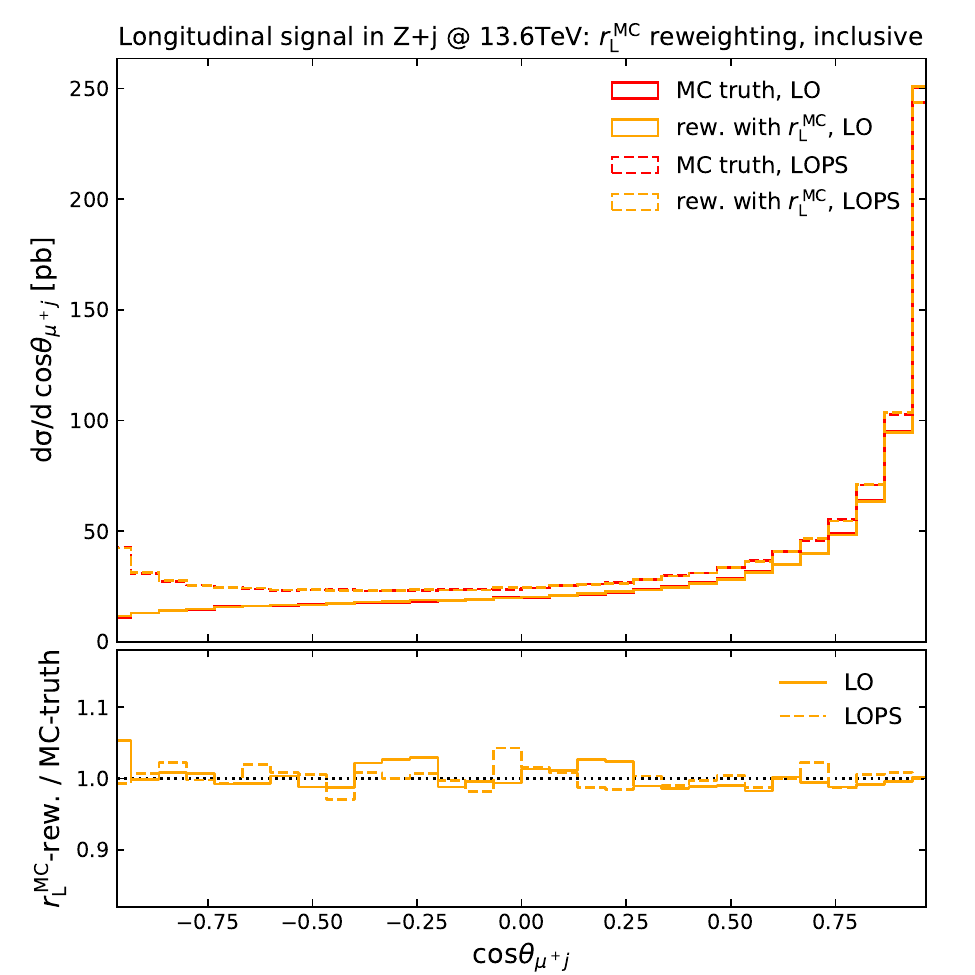}
        \end{subfigure}
        \hfill
        \begin{subfigure}{0.49\textwidth}
%                 \subcaption{}
                 \includegraphics[width=\textwidth,page=3]{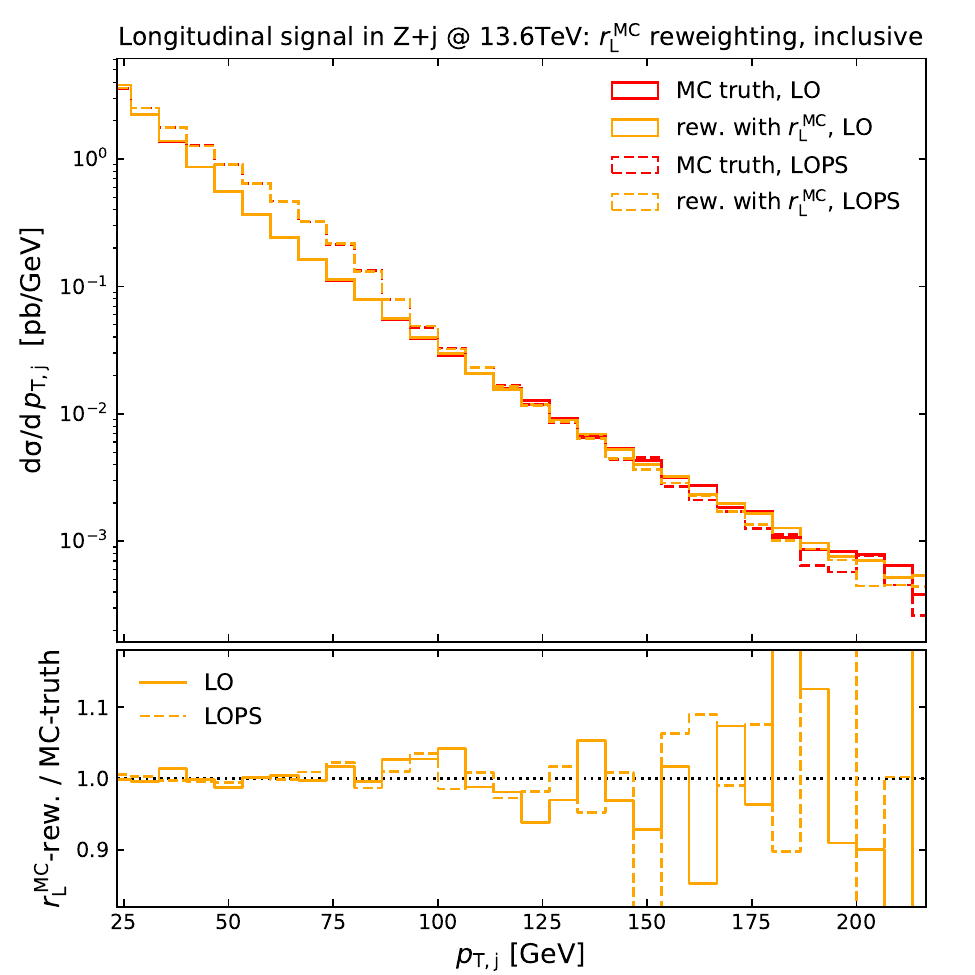}
        \end{subfigure}        
        \caption{
        \label{fig:MCtruth_vs_rLtruthRew}
        Longitudinal reweighting of unpolarised events with $r_{\rL}$ (orange) compared to MC-truth longitudinal events (red) at LO (solid) and LO+PS (dashed). Absolute differential cross sections are shown in the top panel, ratios of reweighted results over MC-truth ones are shown in the bottom panel.
        %The following observables are considered: rapidity of the muon (left), invariant mass of antimuon-jet system (right). 
        The following observables are considered: cosine of the angular separation between the antimuon and the leading jet (left), leading-jet transverse momentum (right).   
        The inclusive setup is understood here. 
        }
\end{figure}

\paragraph{$r_\rL$-reweighting/tagging}
The first key observation that we have made above is that unpolarised event samples can be reweighted/tagged\footnote{In the following, we use indistinguishably \emph{reweighting} and \emph{tagging} unless otherwise stated. In the reweighting approach, the event weights $r_\rL$ (or $\tilde r_\rL$) are directly used to compute longitudinal distributions. In the tagging approach, the longitudinal sample is extracted from the unpolarised one by means a one-dimensional sampling according to $r_\rL$ (or $\tilde r_\rL$) weights. The selected events are then used to compute the longitudinal distributions. Notice that the two methods are equivalent within statistical uncertainties.} using $r_\rL$ to obtain longitudinally polarised samples.
This statement does not only hold at LO but also when including PS effects, thanks to the factorisation of the radiative corrections as implemented in a PS, \ie adding multiple QCD and QED radiations in the collinear approximation at leading-logarithmic accuracy.
This can be seen in Fig.~\ref{fig:MCtruth_vs_rLtruthRew}, where two differential distributions are shown at LO and LO+PS accuracy. The distributions obtained with the $r_\rL$-reweighting reproduce very well those obtained with longitudinal events generated with the Monte Carlo (MC truth).
This confirms that a longitudinally polarised sample can be obtained by simply reweighting an unpolarised one with $r_\rL$ factors.
In addition, one can observe that while the PS corrections are sizeable (boh in the overall normalisation and in the distribution shapes),
the statement about the reweighting is equally true in the presence of PS corrections.
From the results of Table~\ref{tab:validation}, it can also be appreciated how the  $r_\rL$-reweighting performs well both in inclusive setups and in the presence of more exclusive selection cuts. 
\begin{table}[t]
  \begin{center}
    \begin{tabular}{lcc}%
     % \hline %
      accuracy  & MC truth  & $r_\rL$-reweighting\\
      %\hline
     \cellcolor{blue!14} &  \multicolumn{2}{c} {\cellcolor{blue!14} inclusive setup}\\  
      %\hline
      LO                & 0.1704(4) & 0.1703(3) \\
      LO + PS  & 0.1722(4) & 0.1725(3) \\
      %\hline
     \cellcolor{blue!14} & \multicolumn{2}{c} {\cellcolor{blue!14} fiducial setup}\\
      %\hline
      LO                & 0.1879(6) & 0.1883(6) \\
      LO + PS & 0.1889(6) & 0.1894(6) \\
    \end{tabular}
  \end{center}
  \caption{
  Longitudinal-polarisation fraction determined from MC-truth longitudinal events and from $r_\rL$-reweighting of unpolarised events, in the inclusive and fiducial setups.
  Monte Carlo uncertainties on the fractions are shown in parentheses.
  }
  \label{tab:validation}
\end{table}
The differential results analogous to those of Fig.~\ref{fig:MCtruth_vs_rLtruthRew} but in the fiducial setup (not shown here) also highlight an almost perfect behaviour of the reweighting method as in the inclusive setup.

\paragraph{Leading order}
Turning the problem around, the results detailed in the previous paragraph imply that experimental data (here idealised by LO+PS unpolarised events) can be used to extract polarisation fraction provided that $r_\rL$ is known and can be computed on an event-by-event basis.
Actually, $r_\rL$ cannot be computed from experimental data, which do not give access to the full kinematic dependence (including the initial state) and to the flavour of all external particles.
To bypass this issue, one can use NNs to obtain an approximation $\tilde r_\rL$ of the true ratio, based on an incomplete information, namely the one available experimentally which consists in the visible final-state momenta.
Along this line, the first step is therefore to check if one can obtain a good approximation of $r_\rL$ at LO by training a \NN in a supervised setting, as described above, with $r_\rL$ as input label and the final-sate jet momentum and lepton momenta as incomplete information for the training features.

\begin{table}[t]
  \begin{center}
    \begin{tabular}{lccccc}%
     % \hline %
     \cellcolor{blue!14} setup  & \cellcolor{blue!14} MC & \cellcolor{blue!14} \NN\!$_1$ & \cellcolor{blue!14} \NN\!$_1$/MC & \cellcolor{blue!14} \NN\!$_2$ & \cellcolor{blue!14} \NN\!$_2$/MC\\
      %\hline
      inclusive       & 0.1704(4) & 0.1705(5) & 100.1\% & 0.1649(5) & 96.8\%\\
      fiducial      & 0.1879(6) & 0.1904(8) & 101.3\% & 0.1812(8) & 96.4\% \\
      \end{tabular}
  \end{center}
  \caption{
  Longitudinal-polarisation fractions at LO determined from Monte Carlo-truth longitudinal events (MC) and from reweighting of unpolarised events with \NN-predicted $\tilde{r}_\rL$ (\NN), in the inclusive and fiducial setups.
  Monte Carlo uncertainties on the fractions are shown in parentheses. The \NN-predicted fractions are assigned Monte--Carlo-like uncertainties according to the number of events at testing level. 
  }
  \label{tab:fixedorder}
\end{table}

\begin{figure}
        \setlength{\parskip}{-10pt}
        \begin{subfigure}{0.49\textwidth}
%                 \subcaption{}
                 \includegraphics[width=\textwidth,page=9]{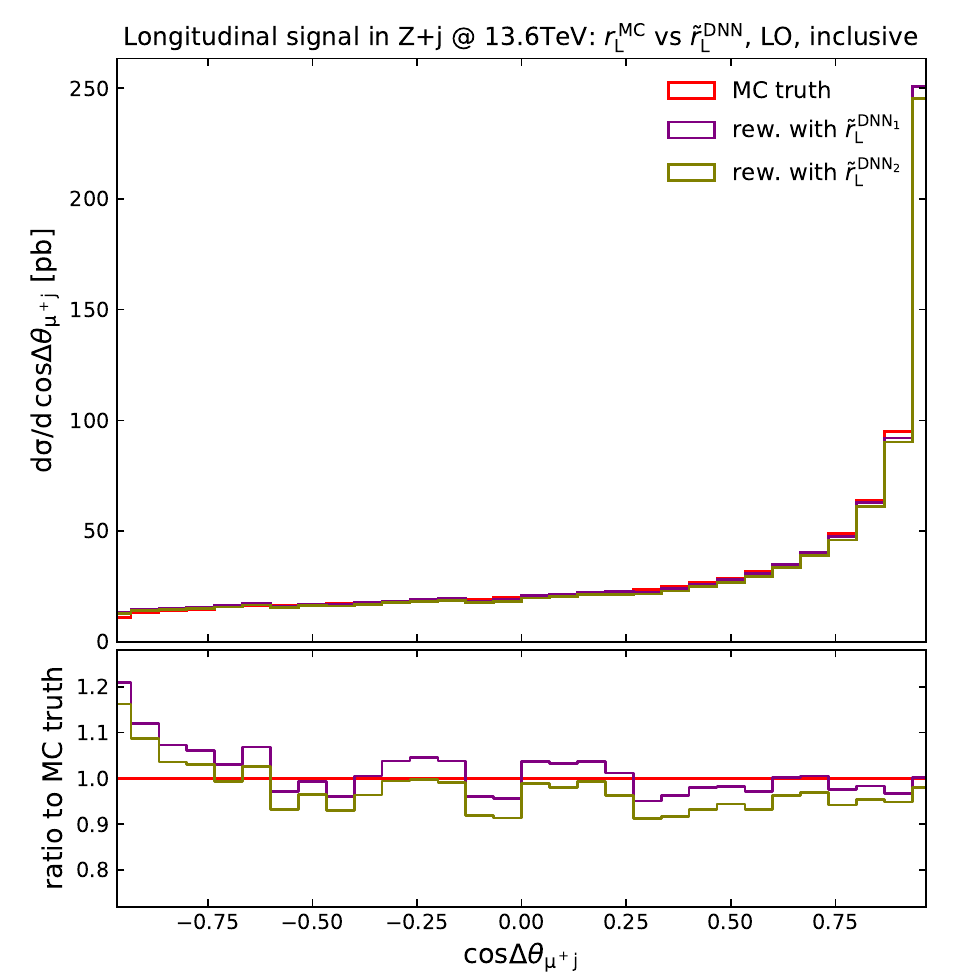}
        \end{subfigure}
        \hfill
        \begin{subfigure}{0.49\textwidth}
%                 \subcaption{}
                 \includegraphics[width=\textwidth,page=3]{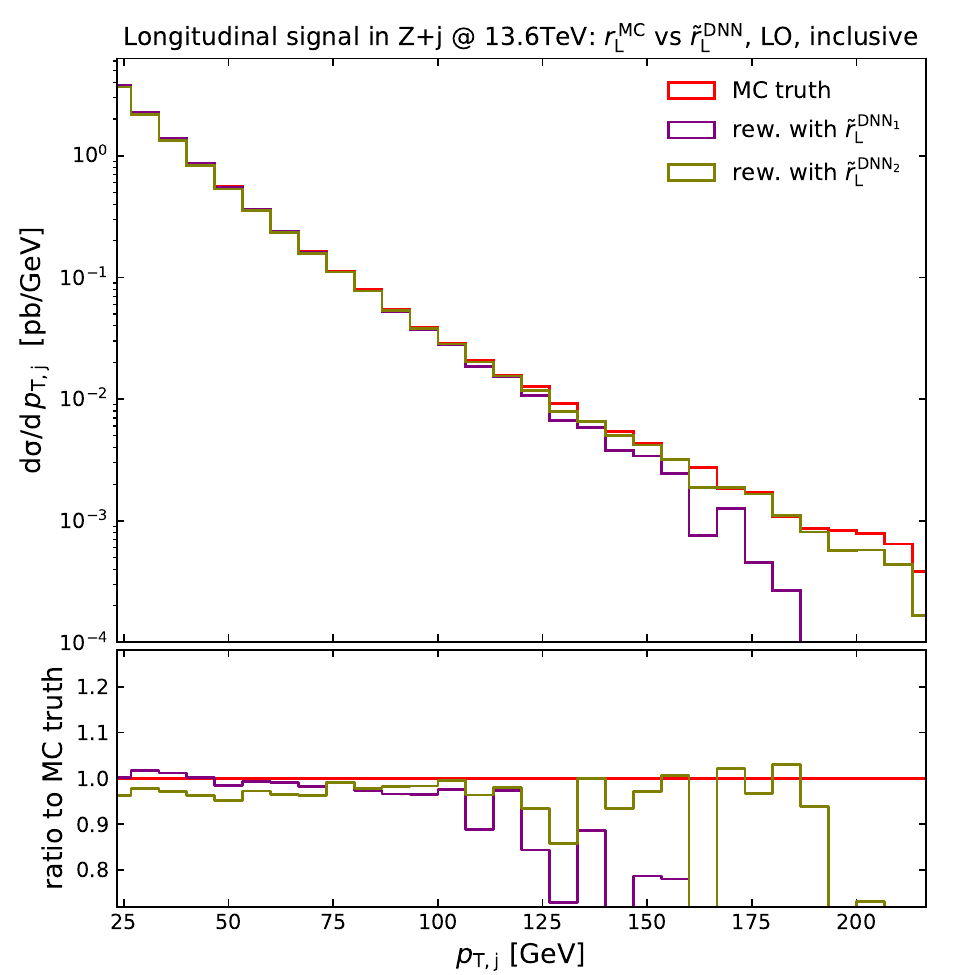}
        \end{subfigure}        
        \caption{
        \label{fig:MCtruth_vs_rLtildeDNN_LO_inc}
        Longitudinal reweighting of unpolarised events with $\tilde{r}_{\rL}$ predicted by two different \NN models (olive and purple curves) compared to MC-truth longitudinal events (red curve) at LO. Absolute differential cross sections are shown in the top panel, ratios of reweighted results over MC-truth ones are shown in the bottom panel. The following observables are considered: cosine of the angular separation between the antimuon and the leading jet (left), leading-jet transverse momentum (right). The inclusive setup is understood. 
        }
\end{figure}
\begin{figure}
        \setlength{\parskip}{-10pt}
        \begin{subfigure}{0.49\textwidth}
%                 \subcaption{}
                 \includegraphics[width=\textwidth,page=9]{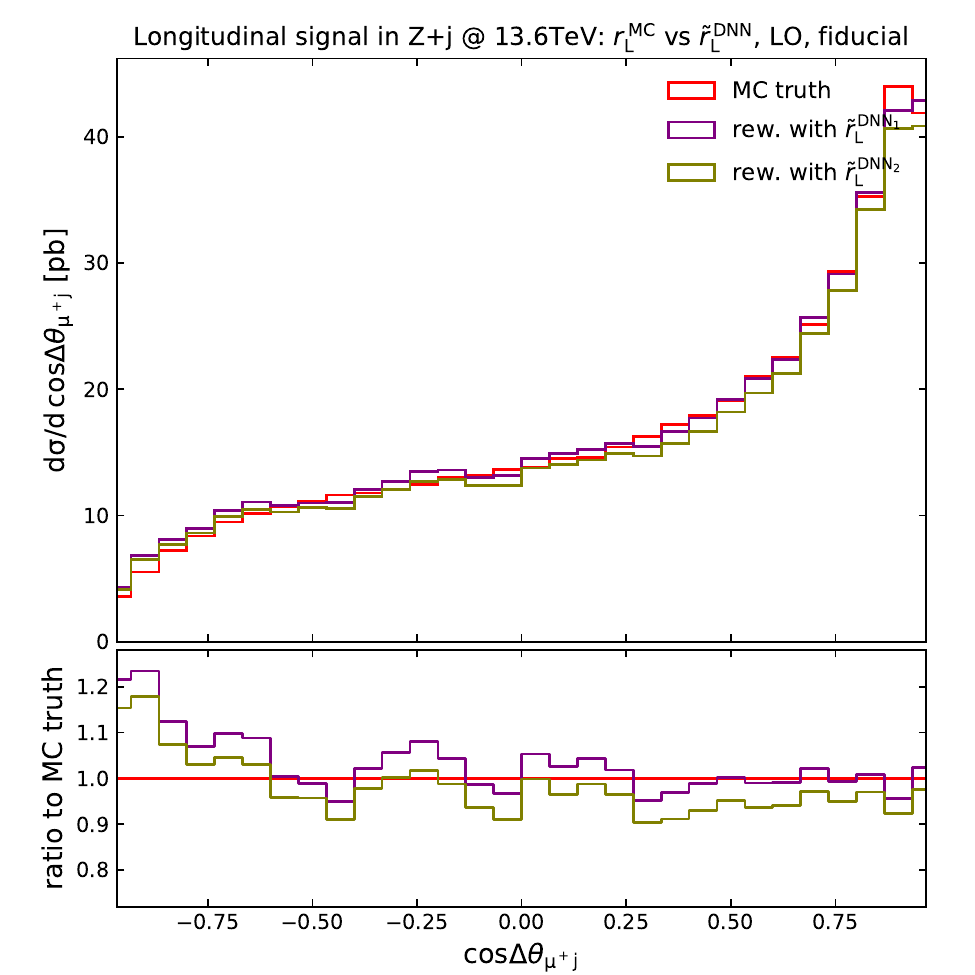}
        \end{subfigure}
        \hfill
        \begin{subfigure}{0.49\textwidth}
%                 \subcaption{}
                 \includegraphics[width=\textwidth,page=3]{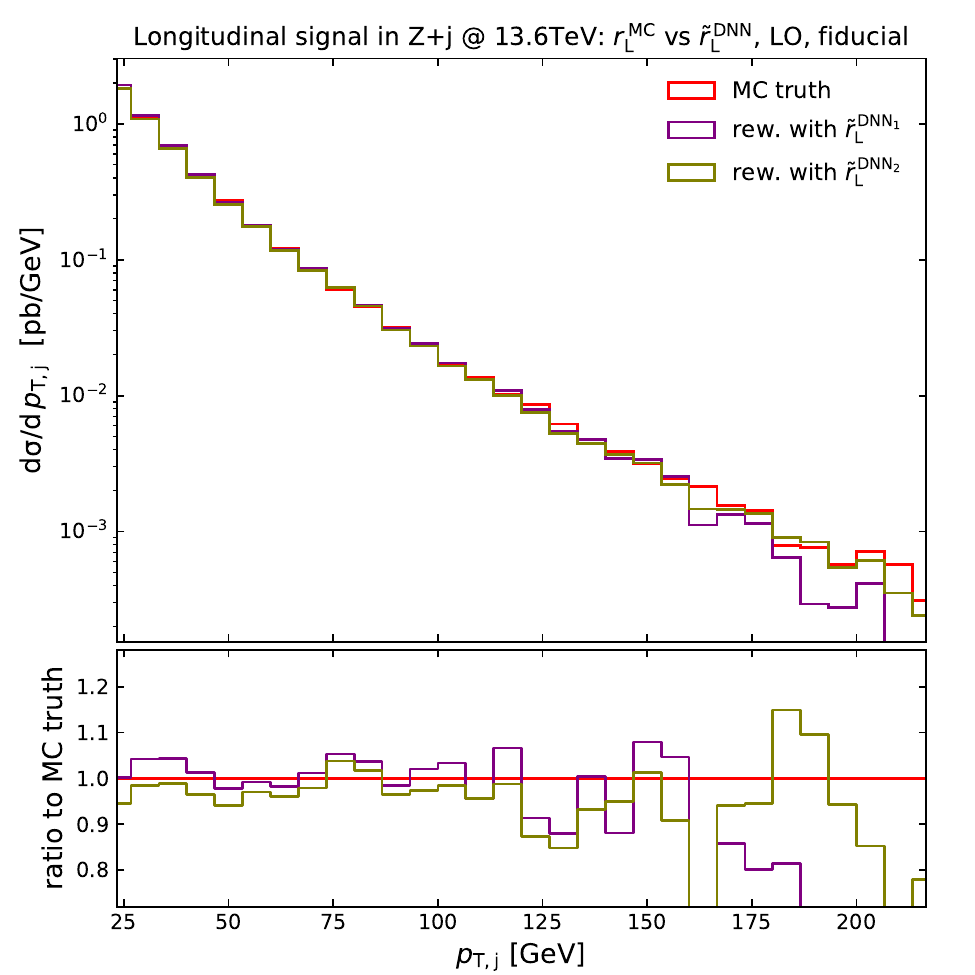}
        \end{subfigure}        
        \caption{
        \label{fig:MCtruth_vs_rLtildeDNN_LO_fid}
        Same structure as Fig.~\ref{fig:MCtruth_vs_rLtildeDNN_LO_inc}. The fiducial setup is understood.
        }
\end{figure}

In Table~\ref{tab:fixedorder} and Figs.~\ref{fig:MCtruth_vs_rLtildeDNN_LO_inc}--\ref{fig:MCtruth_vs_rLtildeDNN_LO_fid}, two different NNs, predicting $\tilde{r}_\rL$ factors as approximations of the $r_\rL$ ones, are compared against the true Monte Carlo results, both at the level of polarisation fractions and at the level of differential cross sections.
The first network (labeled \NN$\!\!_{1}$, purple curves in plots) underwent training for $10^6$ epochs, employing a batch size of 100 and a learning rate of $10^{-4}$. The second network (labeled \NN$\!\!_{2}$, olive curves in plots) was trained for $2\times 10^3$ epochs, using a batch size of $500$ and the same learning rate.
The {generation-level} events were used for the training.

The results of Table~\ref{tab:fixedorder} show a good performance of the \NN\!\!$_{1}$ model in reproducing the polarisation fractions both at inclusive and fiducial level with sub-per-cent accuracy. The \NN$\!\!_{2}$ model underestimates the longitudinal fraction by 3-4\%.

In Figs.~\ref{fig:MCtruth_vs_rLtildeDNN_LO_inc} and \ref{fig:MCtruth_vs_rLtildeDNN_LO_fid}, the cosine of the angular difference between the positive lepton and the jet as well as the transverse momentum of the jet are shown.
The two figures differ in their phase-space regions: Fig.~\ref{fig:MCtruth_vs_rLtildeDNN_LO_inc} is for the inclusive setup while Fig.~\ref{fig:MCtruth_vs_rLtildeDNN_LO_fid} is for the fiducial one.
One observes that the first \NN is reproducing better the true result.
In general, the agreement is at the per-cent level for the phenomenologically relevant part of the phase space and therefore good enough for our purpose.
Also, it is worth pointing out that in suppressed regions of phase space where the statistics is low, the agreement degrades substantially.
\change{
  The limited statistics used for the training stage in this suppressed region
  does not constrain strongly enough the NN model, leading therefore to a systematic error in the NN-model prediction for $\tilde{r}_\rL$.
}
For example, above $150\GeV$ for the transverse momentum of the jet in Fig.~\ref{fig:MCtruth_vs_rLtildeDNN_LO_fid}, the agreement is worth than $20\%$.
This is nonetheless not an issue given that this region is suppressed by two orders of magnitude, meaning that it contributes to about $1\%$ to the cross section and therefore introduces only a per-mille error or less in total.

Finally, we note that the results for the inclusive and fiducial setups are equally good.
The only difference that one can notice is that the fiducial results suffer from larger fluctuations.
This can be attributed to the lower statistics used in the fiducial case ($\approx 220$k events), owing to more restrictive selections that cut away more than half of unpolarised events used in the inclusive setup ($\approx 480$k events).

\paragraph{Parton-shower effects}
Overall, the above results prove that the method is reliable also in typical experimental regions.
Nonetheless, this is a simplified version of the problem as this exercise was performed at LO meaning for events of identical multiplicity.
A more realistic description of the data necessarily requires PS corrections.
Indeed LHC events are typically affected by several effects such as multi-particle interactions, beam remnants, hadronisation, extra QCD and QED radiations etc.
These phenomena are well described by multi-purpose PS programs like {\sc Pythia} \cite{Sjostrand:2014zea}.
In our case, we have included QCD and QED radiations but other effects could equally be included.

\begin{figure}
        \setlength{\parskip}{-10pt}
        \begin{subfigure}{0.49\textwidth}
%                 \subcaption{}
                 \includegraphics[width=\textwidth,page=9]{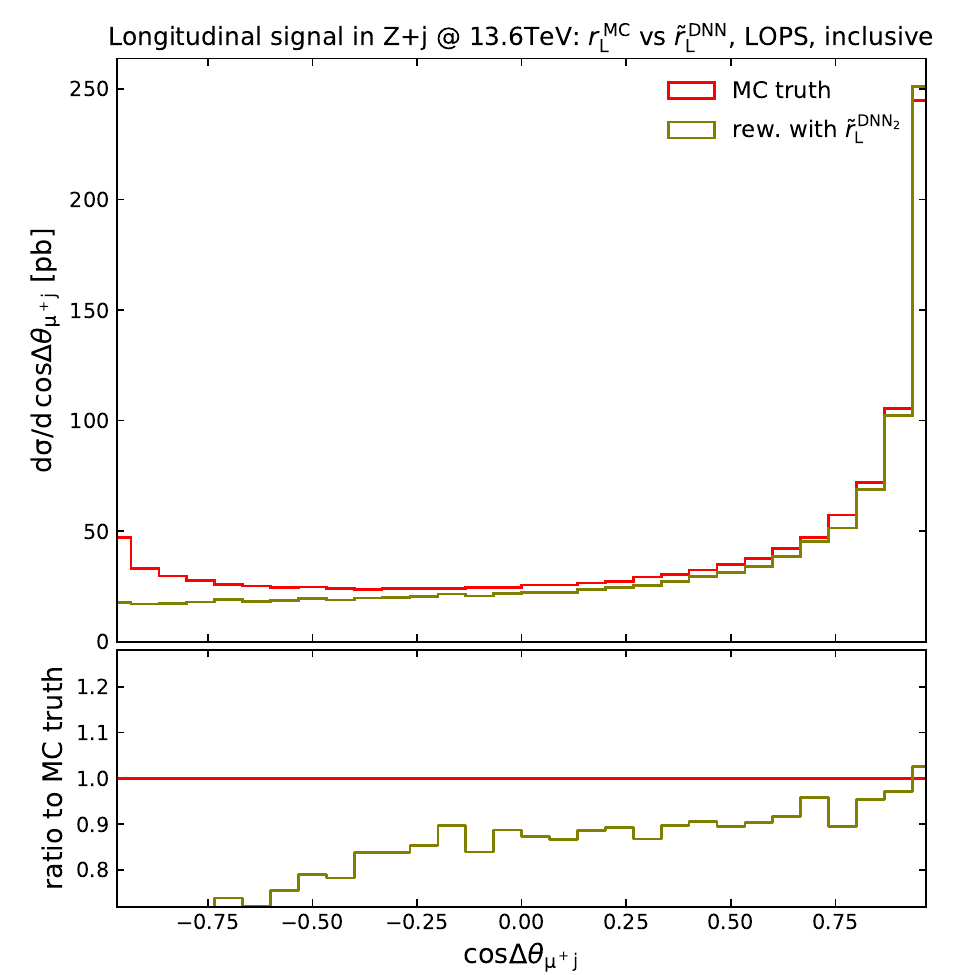}
        \end{subfigure}
        \hfill
        \begin{subfigure}{0.49\textwidth}
%                 \subcaption{}
                 \includegraphics[width=\textwidth,page=3]{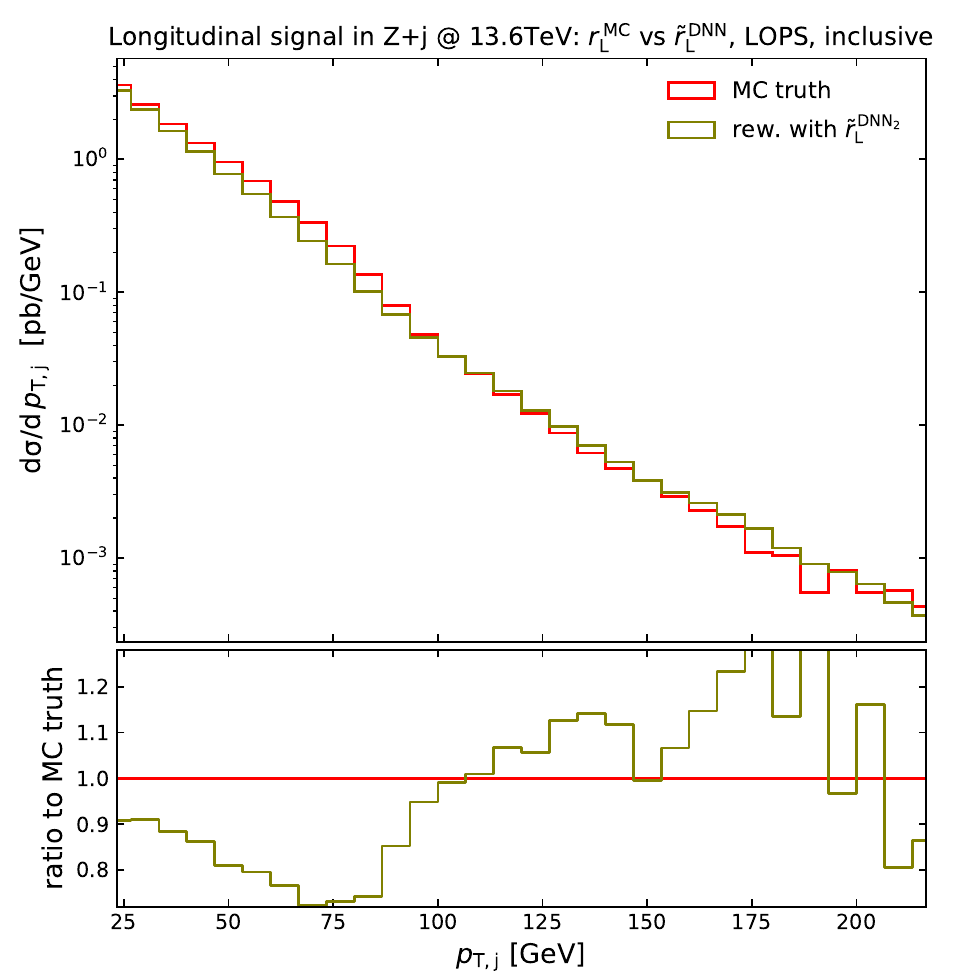}
        \end{subfigure}
        
        \caption{
        \label{fig:DNN_FO} 
            Longitudinal reweighting of unpolarised LO+PS events with $\tilde{r}_{\rL}$ from the \NN model trained with LO events (olive curve), compared to MC-truth longitudinal events (red curve) at LO+PS (QCD shower only). Absolute differential cross sections are shown in the top panel, ratios of reweighted results over MC-truth ones are shown in the bottom panel. 
            The following observables are considered: cosine of the angular separation between the antimuon and the leading jet (left), leading-jet transverse momentum (right). The inclusive setup is understood.
        }
\end{figure}

In order to account for PS effects in our method, one can try to use the previous \NN trained with LO events and apply it to events modeled with PS corrections.
The results of this procedure are shown in Fig.~\ref{fig:DNN_FO} for 
the cosine of the angular separation between the antimuon and the leading jet, and for the leading-jet transverse momentum. Notice that in this case, we have only included effects from QCD PS, avoiding further photon radiations.
From the plots, it is rather clear that this approach is failing.
The reason for this is that the PS generates more QCD radiations leading to a sizeable distortion of the event kinematics. 
In fact, comparing Fig.~\ref{fig:DNN_FO} with the fixed-order results in Fig.~\ref{fig:MCtruth_vs_rLtruthRew}, one observes that applying the \NN trained with LO events to LO+PS events tends to reproduce the LO distribution shapes rather than the LO+PS ones.

As shown previously, one can apply a reweighting of the unpolarised sample and then apply the PS procedure or viceversa in order to obtain a longitudinally polarised sample with PS effects.
This also means that for each showered event one can compute $r_\rL$ with the original LO momenta before PS and therefore associate a meaningful $r_\rL$ to each showered event.
One can therefore train a new \NN with the original $r_\rL$ (computed before showering) along with the momenta after showering.
Given that showered events possess more than one jet, only the four momentum of the jet with the largest transverse momentum is provided as a feature for the \NN-model training.

In order to tackle this problem, we adopted the following training strategy. 
First, a wide neural network (labeled \NN$\!\!_{\rm ws}$) is trained using a \emph{warm-start initialisation} \cite{yosinski2014transferable}. Utilising the warm-start approach in training NNs involves initialising the model with weights from a previously trained model. This strategy potentially accelerates convergence, enhances performance, and reduces the need for extensive data, creating an efficient framework for model training. From a physics intuition, in the generation chain that starts with LO process and move to LO+PS, this step does not represents a completely new learning task rather it is a perturbation of the original process.
It follows that the procedure involved utilising the configuration of the network previously trained on the LO events and starting the training with the LO+PS events from its optimal configuration in terms of architecture of the \NN and its relative weights. This procedure, which is well-known in other machine-learning applications, has not been yet fully exploited in high-energy physics. Skipping a complete \NN architecture optimisation procedure is advantageous because of the faster identification of the best model
and of the computational-resource saving.
For the sake of comparison, a second general \NN (labeled \NN$\!\!_{\rm nows}$) is built from scratch, looking for the best depth-to-width ratio with a randomly chosen initial configuration. 

\begin{table}[t]
  \begin{center}
    \begin{tabular}{lccccc}%
     % \hline %
     \cellcolor{blue!14} setup  & \cellcolor{blue!14} MC & \cellcolor{blue!14} \NN\!$_{\rm ws}$ & \cellcolor{blue!14} \NN\!$_{\rm ws}$/MC & \cellcolor{blue!14} \NN\!$_{\rm no\, ws}$ & \cellcolor{blue!14} \NN\!$_{\rm no\, ws}$/MC\\
      %\hline
      %\hline
      inclusive       & 0.1722(4) & 0.1705(3) & 99.0\% & 0.1646(3) & 95.6\%\\
      fiducial        & 0.1889(6) & 0.1853(5) & 98.1\% & 0.1791(5) & 94.8\% \\
      \end{tabular}
  \end{center}
  \caption{
  Longitudinal-polarisation fractions at LO+PS determined from Monte Carlo-truth longitudinal events (MC) and from reweighting of unpolarised events with \NN-predicted $\tilde{r}_\rL$ (\NN), in the inclusive and fiducial setups.
  Monte Carlo uncertainties on the fractions are shown in parentheses. The \NN-predicted fractions are assigned Monte--Carlo-like uncertainties according to the number of events at testing level.
  }
  \label{tab:partonshower}
\end{table}

\begin{figure}
        \setlength{\parskip}{-10pt}
        \begin{subfigure}{0.49\textwidth}
%                 \subcaption{}
                 \includegraphics[width=\textwidth,page=9]{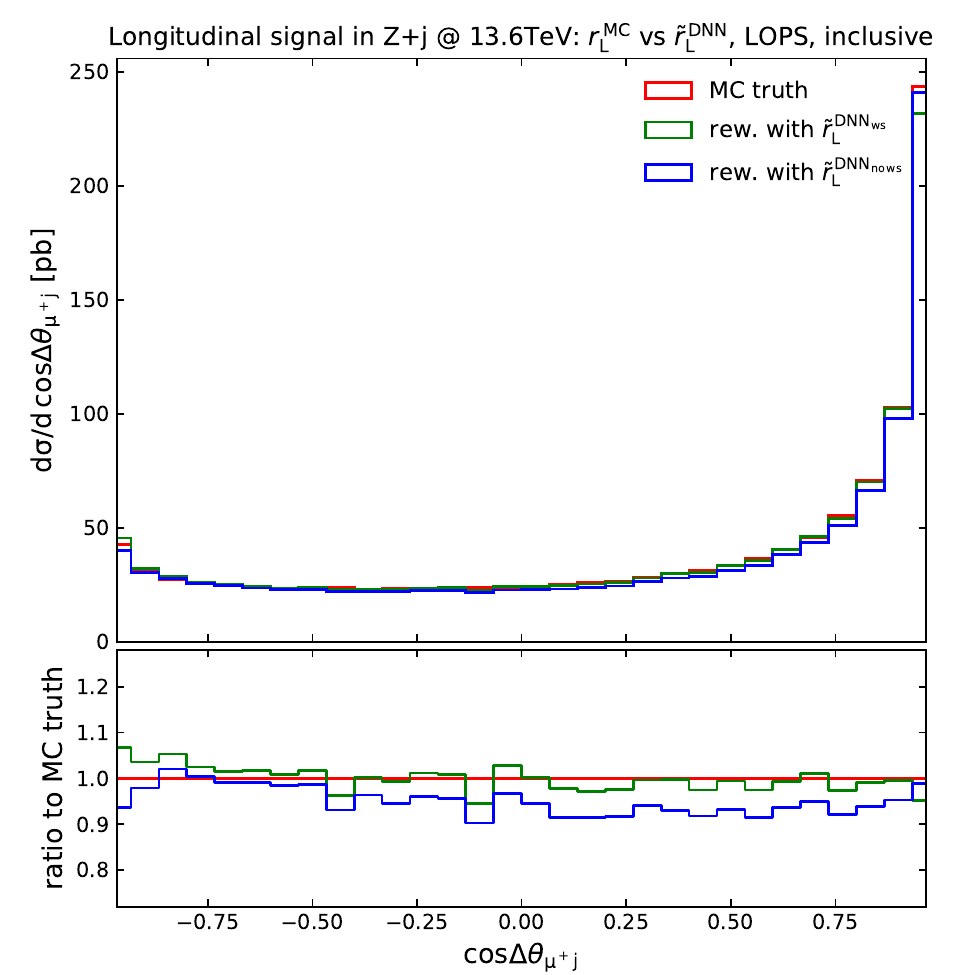}
        \end{subfigure}
        \hfill
        \begin{subfigure}{0.49\textwidth}
%                 \subcaption{}
                 \includegraphics[width=\textwidth,page=3]{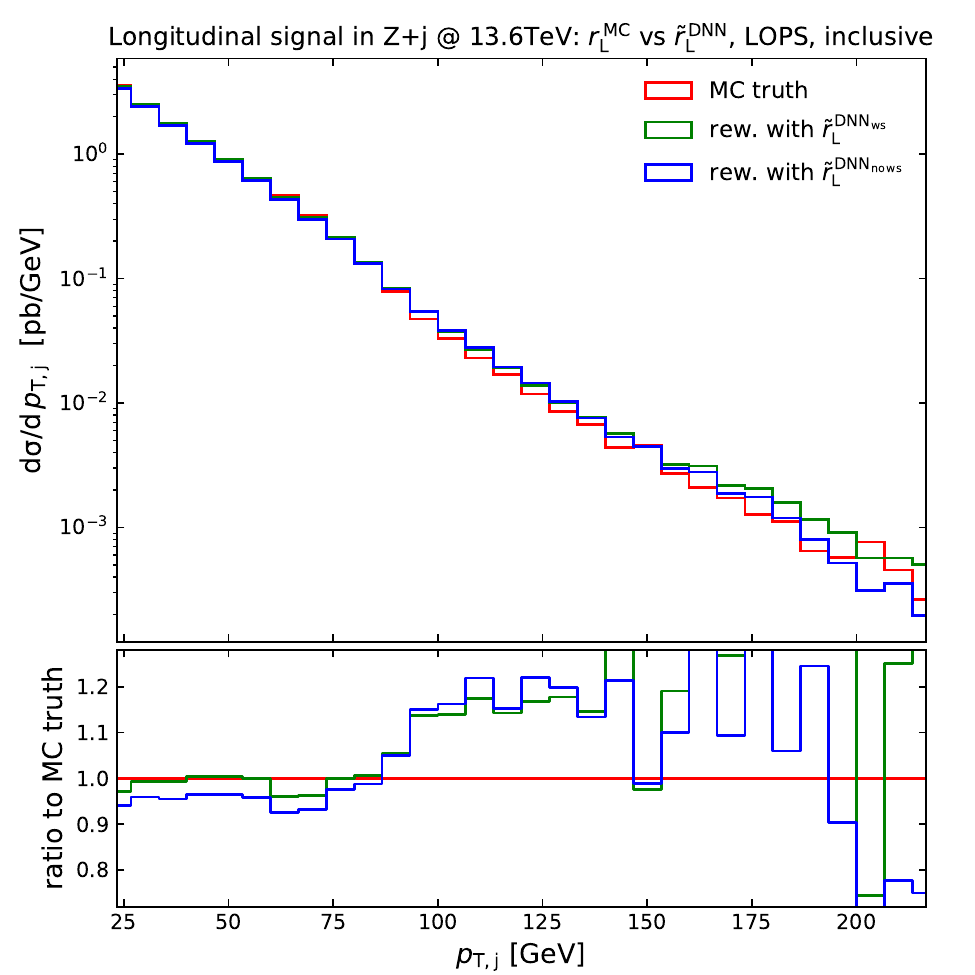}
        \end{subfigure}        
    \caption{
        \label{fig:MCtruth_vs_rLtildeDNN_LO+PS_inc}
        Longitudinal reweighting of unpolarised events with $\tilde{r}_{\rL}$ from two different \NN models (blue and green curves) compared to MC-truth longitudinal events (red curve) at LO+PS (both QCD and QED showers included). Absolute differential cross sections are shown in the top panel, ratios of reweighted results over MC-truth ones are shown in the bottom panel. 
        The following observables are considered: cosine of the angle between the antimuon and the leading jet (left), transverse momentum of the leading jet (right).
        The inclusive setup is understood.
        }
\end{figure}

\begin{figure}
        \setlength{\parskip}{-10pt}
        \begin{subfigure}{0.49\textwidth}
%                 \subcaption{}
                 \includegraphics[width=\textwidth,page=9]{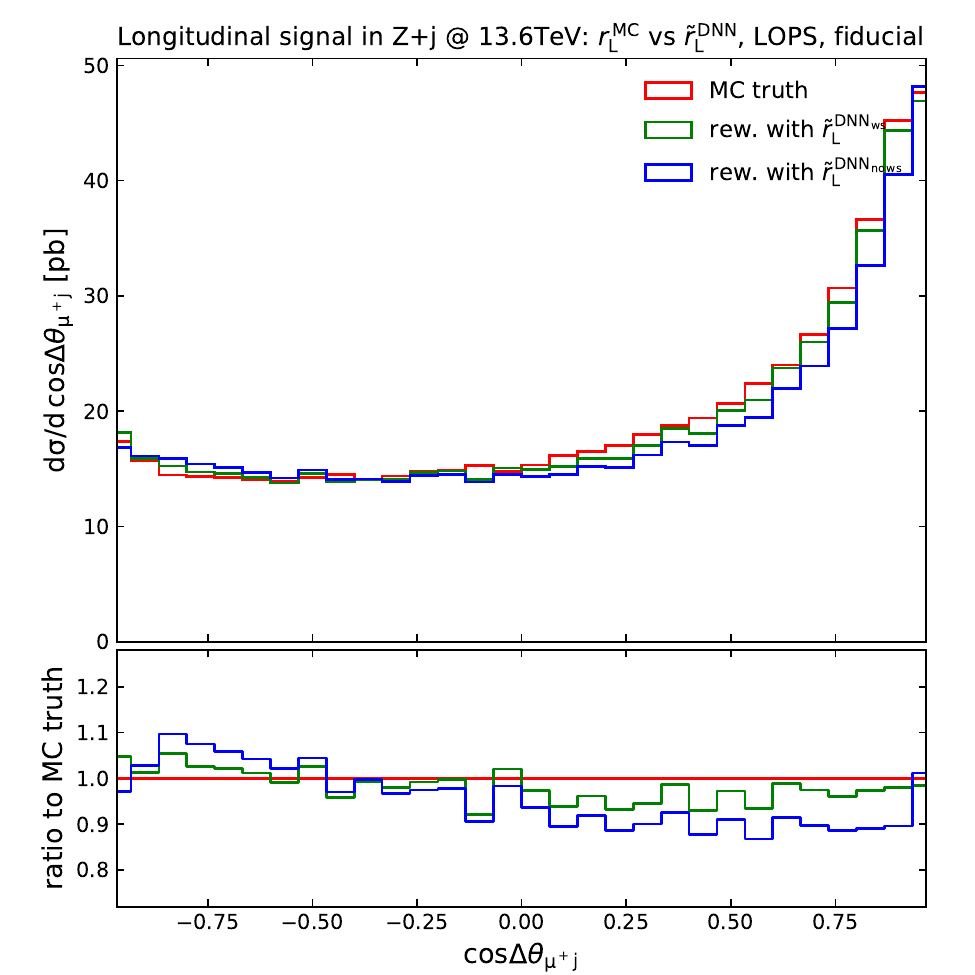}
        \end{subfigure}
        \hfill
        \begin{subfigure}{0.49\textwidth}
%                 \subcaption{}
                 \includegraphics[width=\textwidth,page=3]{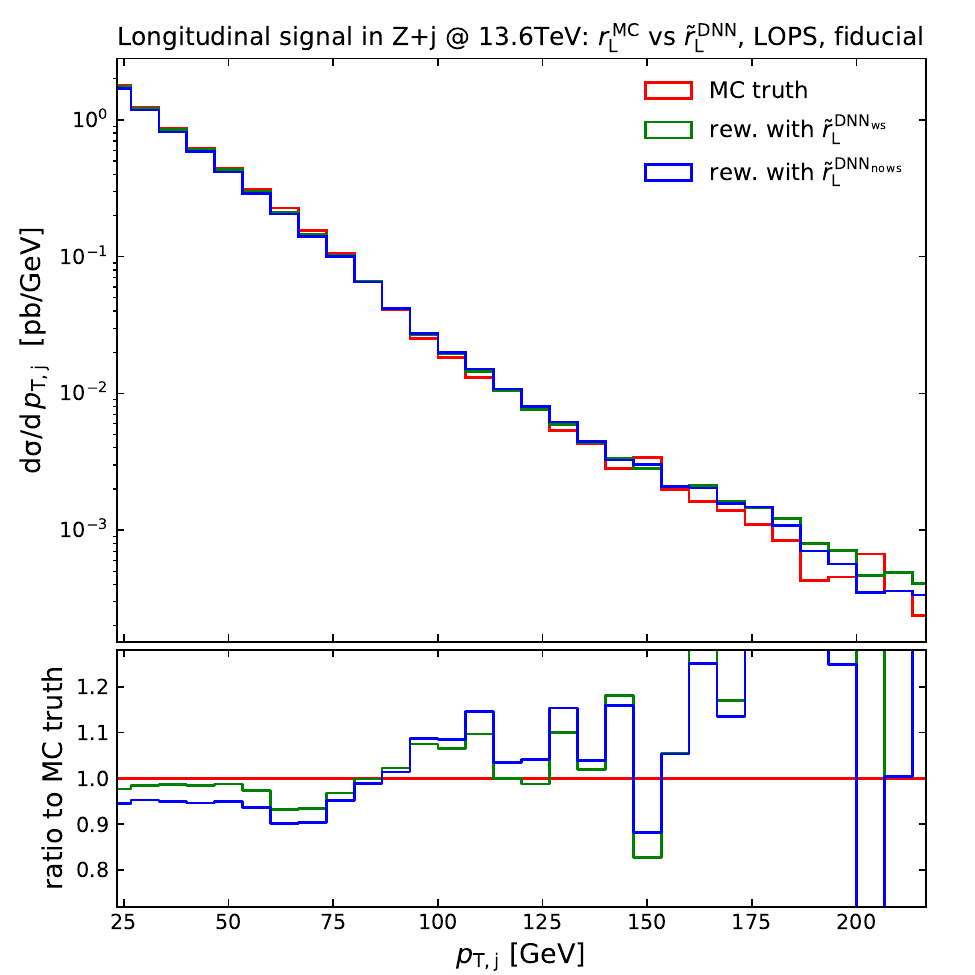}
        \end{subfigure}  
        \caption{
        \label{fig:MCtruth_vs_rLtildeDNN_LO+PS_fid}
        Same structure as Fig.~\ref{fig:MCtruth_vs_rLtildeDNN_LO+PS_inc}. The fiducial setup is understood. 
        }
\end{figure}

The results provided by these two models are reported in Table~\ref{tab:partonshower} and Figs.~\ref{fig:MCtruth_vs_rLtildeDNN_LO+PS_inc}--\ref{fig:MCtruth_vs_rLtildeDNN_LO+PS_fid} (green curves for \NN$\!\!_{\rm ws}$, blue curves for \NN$\!\!_{\rm no\, ws}$).
As one can see from both integrated and differential results that the \NN with warm start is outperforms the one built from scratch.
From the results, it is clear that the warm start has beneficial effects on the NNs. Firstly, it biases the training towards solving a similar task, allowing the network to adjust its parameters to the new data, which limits the search space and leads to faster convergence. Additionally, as the problem becomes easier to solve, the quality of the solution improves. 
From a physics viewpoint, the good behaviour of the \NN$\!\!_{\rm ws}$ implies that the LO step is actually of crucial importance to be able to use this method in an experimental analysis.
In particular, the results are per-cent accurate at the level of polarisation fractions, which is good enough for the level of precision of this study.  
Considering differential observables, Fig.~\ref{fig:MCtruth_vs_rLtildeDNN_LO+PS_inc} refers to the inclusive case while Fig.~\ref{fig:MCtruth_vs_rLtildeDNN_LO+PS_fid} refers to the fiducial case.
The same conclusions as at fixed order hold, namely that the limited statistics do play a role in the accuracy of the method, as can be observed in the \change{far} tails of the transverse-momentum distribution in Fig.~\ref{fig:MCtruth_vs_rLtildeDNN_LO+PS_fid} or in other phase-space regions which are the least populated ones.
\change{
  Nonetheless, at $100\GeV$ in the transverse-momentum distribution of the leading jet, a $10$-$20\%$ mismodelling can be observed.
  These effects cannot be solely attributed to the statistics but should be considered as a systematic error of the NN.
  This mismodelling might originate from PS effect as shown in Fig.~\ref{fig:MCtruth_vs_rLtruthRew} where the region around $100\GeV$
  marks a quantitative change in the PS corrections. This could also be interpreted as a limitation of the NN model to capture all features.
  Nonetheless, these $10$-$20\%$ discrepancies appears in bins that are suppressed by almost two orders of magnitude and therefore
  they are not physically significant when integrating over the whole transverse-momentum spectrum.
  Hence, the method proposed here is still per-cent accurate.
}

\paragraph{Event tagging with $\tilde{r}_\rL$}
As already discussed in Sect.~\ref{sec:sol_prob}, the probabilistic interpretation of ${r}_\rL$ leads to the expectation that the $\tilde{r}_\rL$ predicted by the \NN models is positive. However, the \NN models have no physics insights about this constraint and $\tilde{r}_\rL$ is not always positive.

It turns out that at LO+PS level, the NNs are able to predict positive $\tilde{r}_\rL$ for more than 99\% of the event, both in the inclusive and in the fiducial setup.
Interestingly, at fixed order, the performances are worse, with positive
longitudinal weights predicted for only roughly 95\% of the events.
The events for which the \NN predicts negative weights give a harder $p_{\rm T, \Pj_1}$ spectrum compared to the events with positive $\tilde{r}_\rL$, highlighting that in order to improve the accuracy of the \NN also in boosted regimes, a dedicated training with boosted events is needed.
We checked that discarding the events with negative $\tilde{r}_\rL$, in spite of a partial improvement in the reproduction of the transverse-momentum shapes, overestimates by several per-cent the overall longitudinal fraction. 

A viable strategy could be to include a suppression function for negative $\tilde{r}_\rL$ at the level of the last layer of the \NN models, as a small step toward physics-informed approaches. These ones have already been applied for classification tasks in particle physics, where enforcing symmetries conservation for transformations under the Lorentz group, provides a much more physically interpretable model \cite{bogatskiy2020lorentz}.
However, there is no guarantee of improved accuracy in the \NN. In our specific case, enforcing the positivity of the label actually worsens the overall performance. This is due to the introduction of constraints complicating the landscape of the loss function, resulting in more challenging geometries with multiple local minima. As a result, training becomes less effective, leading to poorer predictions from the model.
We have refrained from investigating this aspect further, as the LO+PS results are satisfactory for the present application.

So far we have indistinguishably used the expression \emph{reweighting} and \emph{tagging}. 
However, while the reweighting strategy can be applied also in the presence of negative weights, the event tagging is not well defined anymore in that case.
In other terms, performing a longitudinal tagging according to the \NN predictions is not possible for events with negative $\tilde r_\rL$. 
While at LO accuracy this means throwing away $5\%$ of the events, at LO+PS accuracy, which is the most important case as it mimics the experimental environment, less than a per cent of the events have to be thrown away, which is good enough for our purposes.

\begin{table}[t]
  \begin{center}
    \begin{tabular}{lccc}%
     % \hline %
     \cellcolor{blue!14} setup  & \cellcolor{blue!14} MC & \cellcolor{blue!14} \NN\!$_{\rm ws}$-sampl & \cellcolor{blue!14} \NN\!$_{\rm ws}$-sampl/MC\\
      %\hline
      %\hline
      inclusive       & 0.1722(4) &  0.1716(5)& 99.7\%\\
      fiducial        & 0.1889(6) &  0.1850(8)& 97.9\%\\
      \end{tabular}
  \end{center}
  \caption{
  Longitudinal-polarisation fractions at LO+PS determined from MC-truth longitudinal events (MC) and sampling (\NN-sampl) of unpolarised events according to $\tilde{r}_\rL$ predicted with the \NN$\!\!_{\rm ws}$ model, in the inclusive and fiducial setups.
  Monte Carlo uncertainties on the fractions are shown in parentheses. The \NN-predicted fractions are assigned Monte--Carlo-like uncertainties according to the number of events at testing level.
  }
  \label{tab:sampling}
\end{table}

\begin{figure}
        \setlength{\parskip}{-10pt}
        \begin{subfigure}{0.49\textwidth}
%                 \subcaption{}
                 \includegraphics[width=\textwidth,page=9]{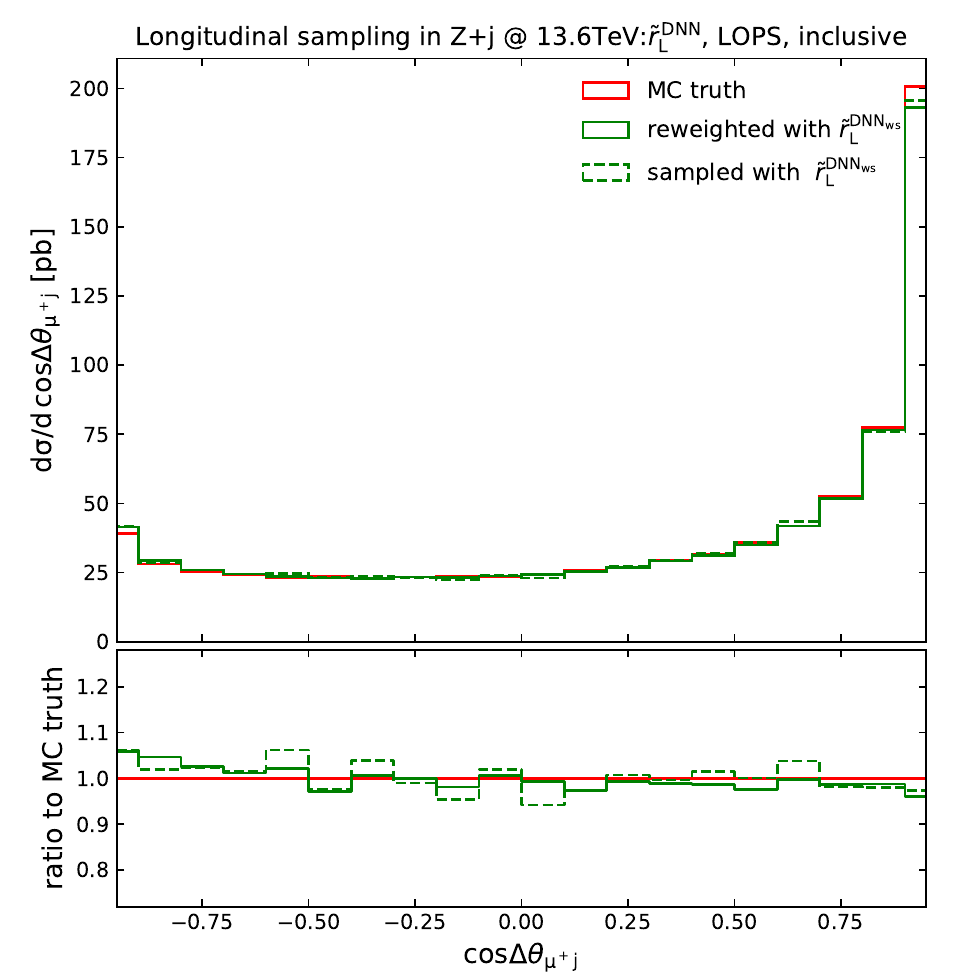}
        \end{subfigure}
        \hfill
        \begin{subfigure}{0.49\textwidth}
                 \subcaption{}
                 \includegraphics[width=\textwidth,page=14]{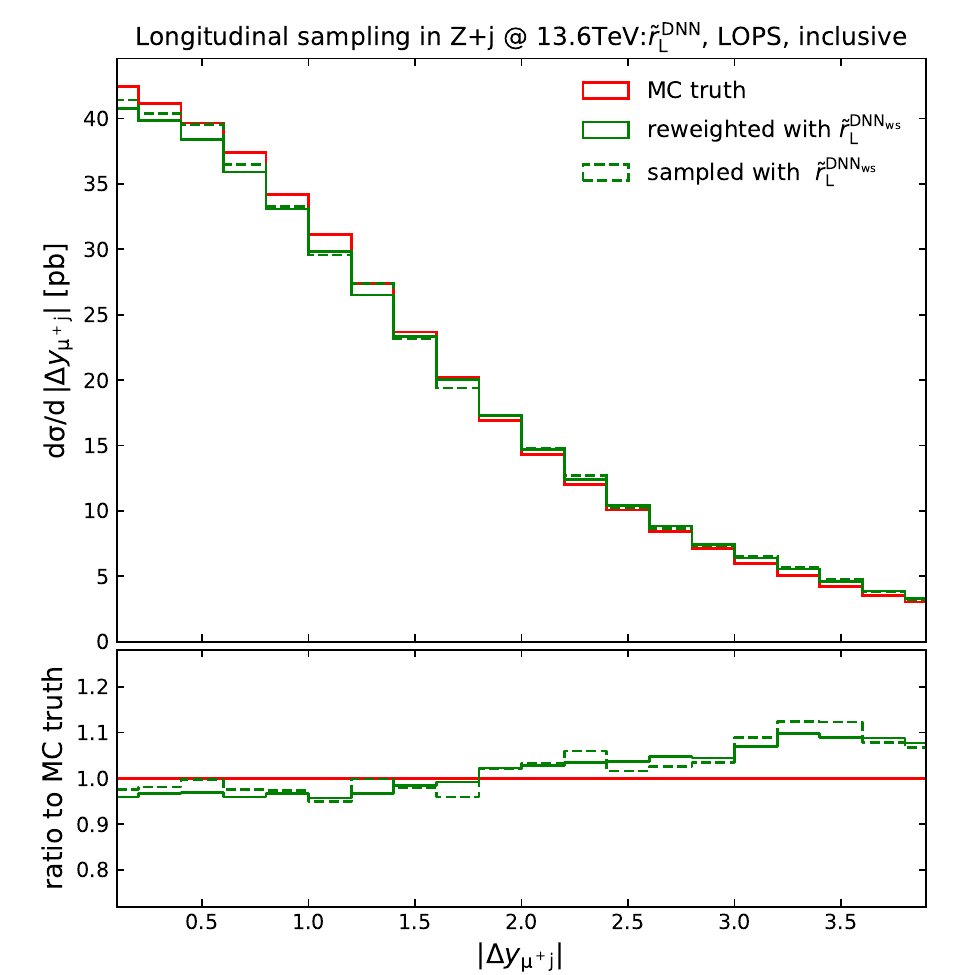}
        \end{subfigure}   
        \begin{subfigure}{0.49\textwidth}
                 \subcaption{}
                 \includegraphics[width=\textwidth,page=7]{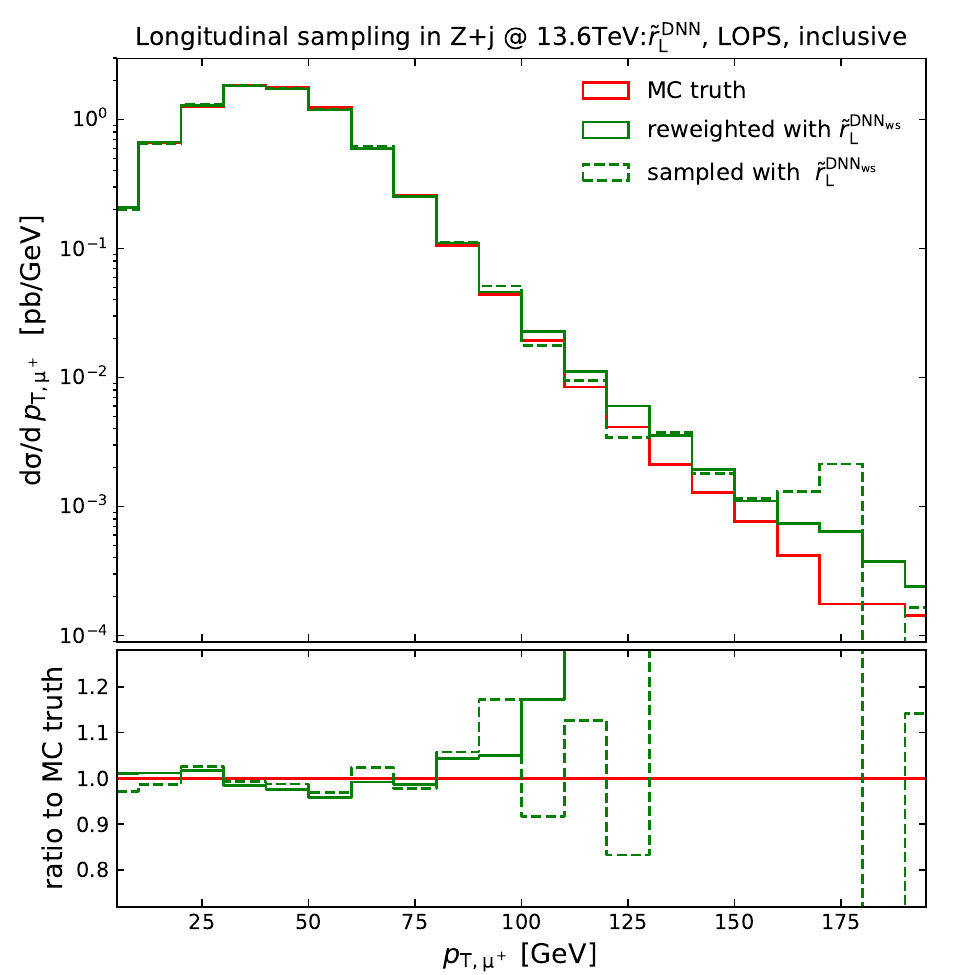}
        \end{subfigure}
        \hfill
        \begin{subfigure}{0.49\textwidth}
                 \subcaption{}
                 \includegraphics[width=\textwidth,page=3]{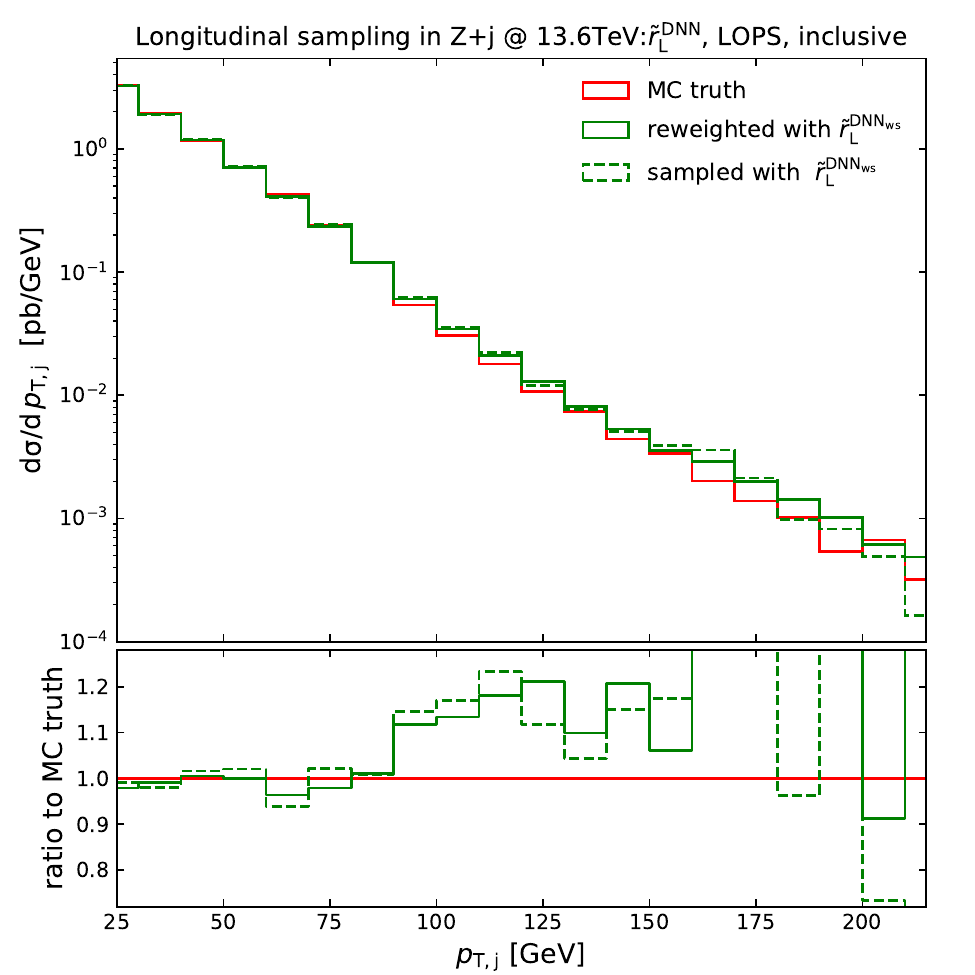}
        \end{subfigure}  
        \caption{
        \label{fig:sampling_LO+PS_inc}
        Longitudinal reweighting (solid green) and tagging (dashed green) of unpolarised events with $\tilde{r}_{\rL}$ predicted by the \NN$\!\!_{\rm ws}$ model compared to MC-truth longitudinal events (solid red) at LO+PS (both QCD and QED showers included). Absolute differential cross sections are shown in top panels, ratios over MC-truth ones are shown in bottom panels. 
        The following observables are considered: cosine of the angle between the antimuon and the leading jet (top left), rapidity separation between the antimuon and the leading jet (top right), tranverse momentum of the antimuon (bottom left), transverse momentum of the leading jet (bottom right).
        The inclusive setup is understood.
        }
\end{figure}

\begin{figure}
        \setlength{\parskip}{-10pt}
        \begin{subfigure}{0.49\textwidth}
%                 \subcaption{}
                 \includegraphics[width=\textwidth,page=9]{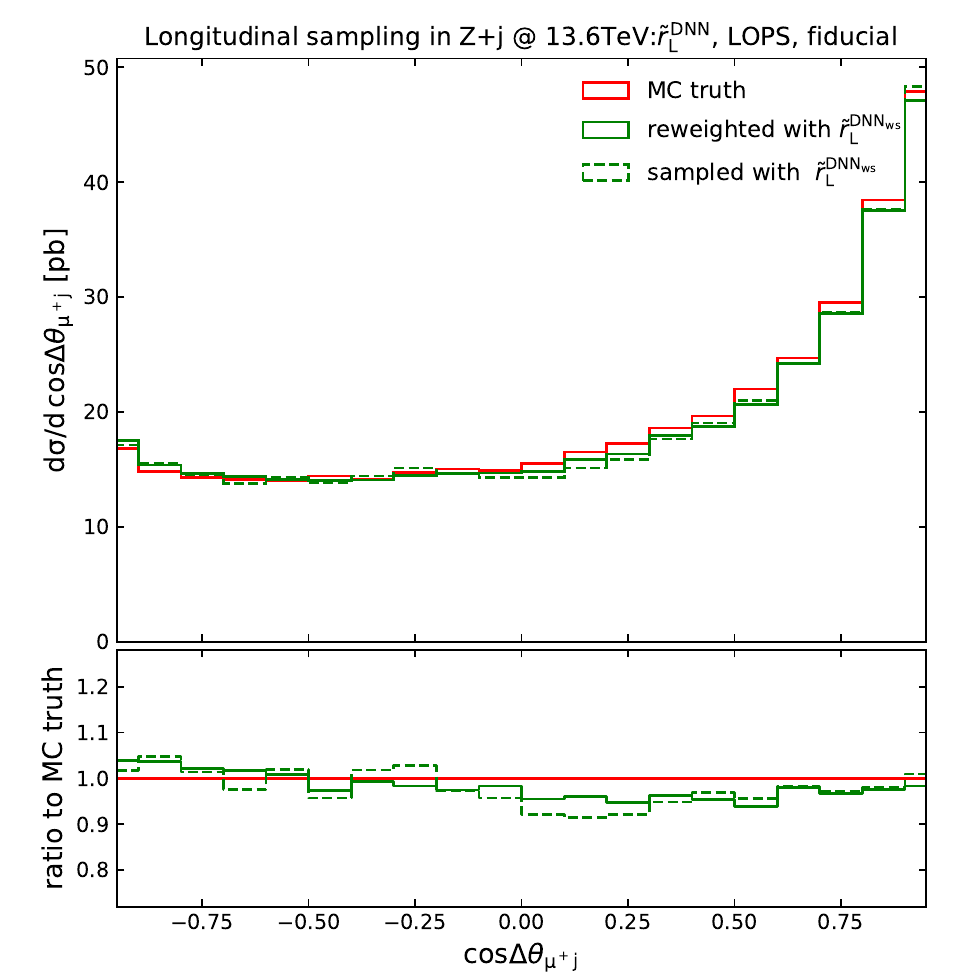}
        \end{subfigure}
        \hfill
        \begin{subfigure}{0.49\textwidth}
                 \subcaption{}
                 \includegraphics[width=\textwidth,page=14]{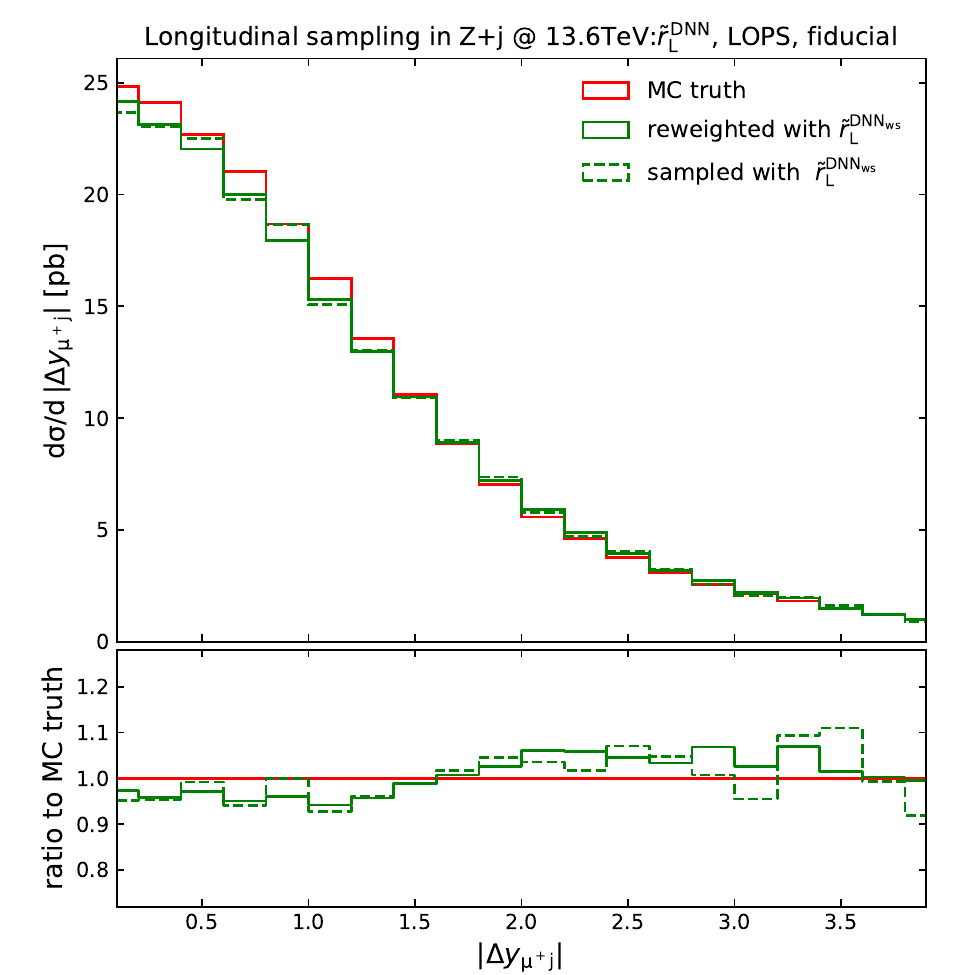}
        \end{subfigure}   
        \begin{subfigure}{0.49\textwidth}
                 \subcaption{}
                 \includegraphics[width=\textwidth,page=7]{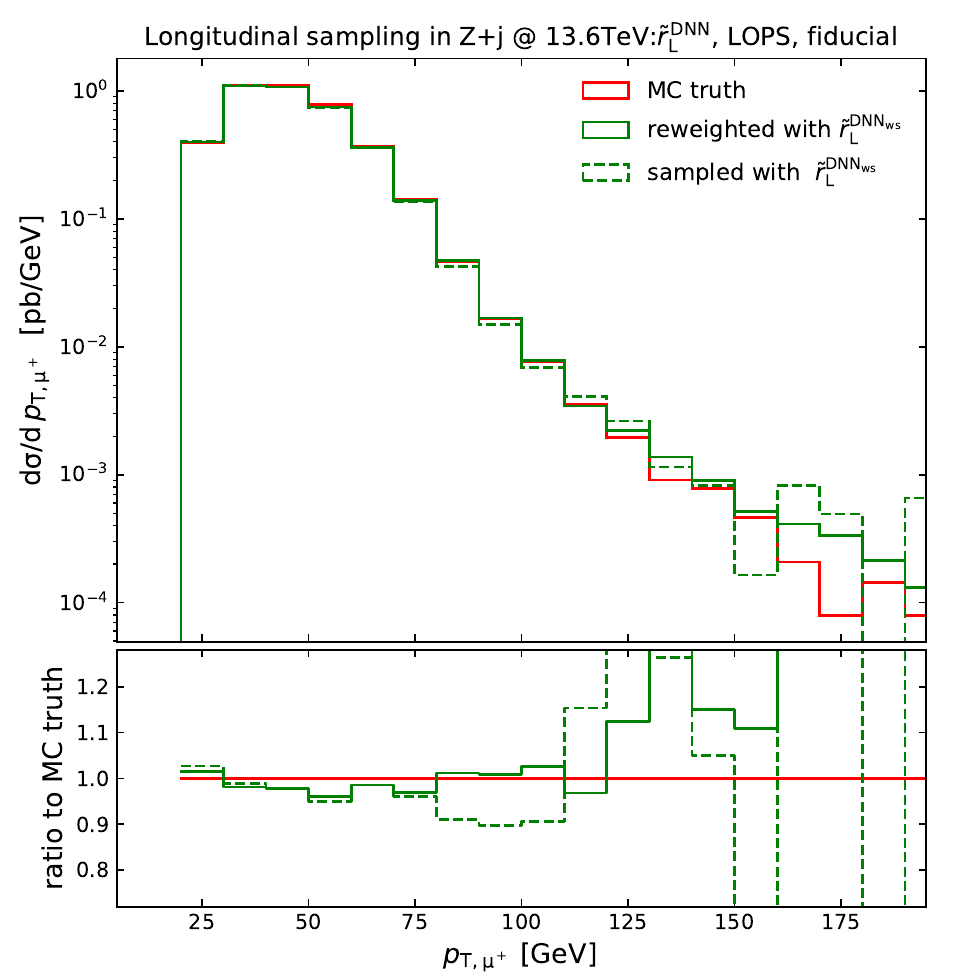}
        \end{subfigure}
        \hfill
        \begin{subfigure}{0.49\textwidth}
                 \subcaption{}
                 \includegraphics[width=\textwidth,page=3]{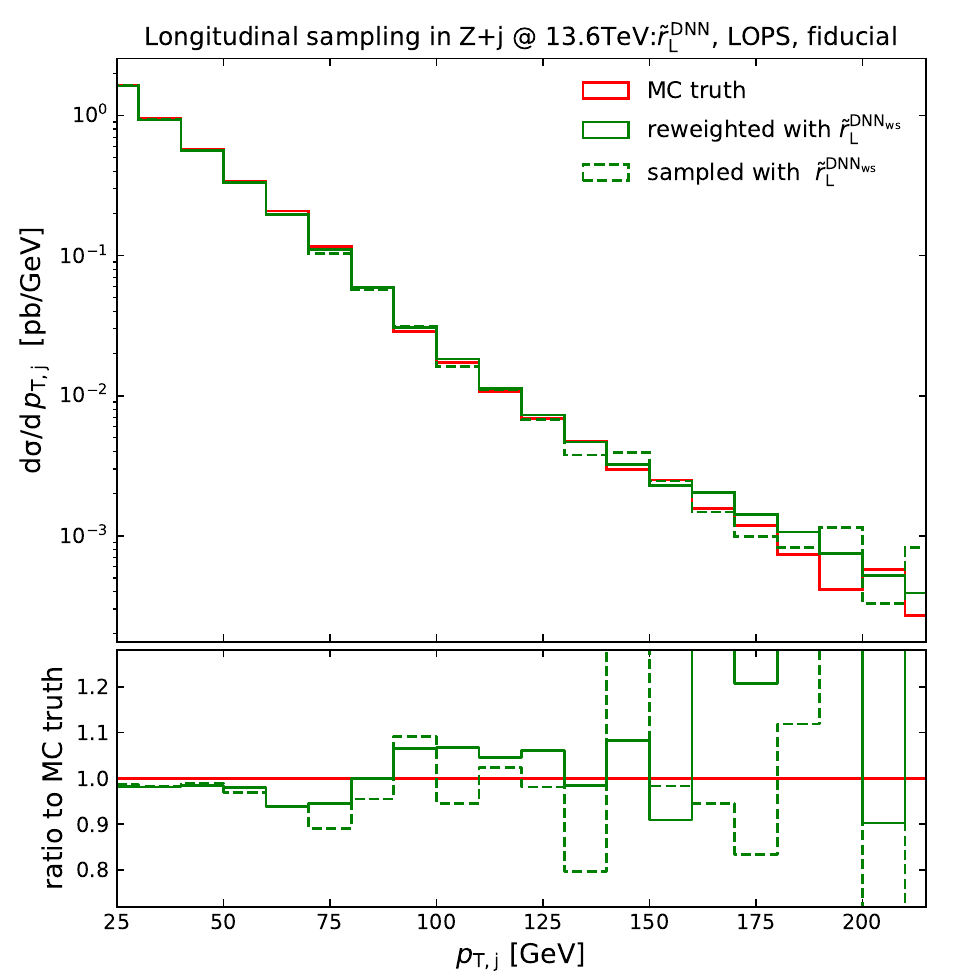}
        \end{subfigure}  
        \caption{
        \label{fig:sampling_LO+PS_fid}
        Same structure as Fig.~\ref{fig:sampling_LO+PS_inc}. The fiducial setup is understood.
        }
\end{figure}

To illustrate the applicability of the method, we show final results for the event tagging, which turns out to be equivalent to the reweighting ones, at LO+PS accuracy.
In Table \ref{tab:sampling}, the polarisation fractions obtained with tagging are compared to the MC-truth ones.
As expected, these results are good and almost equivalent to the reweigthing results provided in Table~\ref{tab:partonshower}, since only very few events have a negative $\tilde r_\rL$.
Notice that for this comparison we have only considered the \NN that employs the warm-start approach (\NN$\!\!_{\rm ws}$).
The results are equally good at the differential level, as can be observed in Figs.~\ref{fig:sampling_LO+PS_inc} and \ref{fig:sampling_LO+PS_fid} for the inclusive and fiducial setup, respectively.
In particular, in these plots, the MC-truth longitudinal distributions are compared with those obtained  reweighting and tagging according to $\tilde r_\rL$.
Both are equivalent up to statistical fluctuations.
This finally demonstrates that using $\tilde r_\rL$ with experimental inputs enables an actual longitudinal-polarisation tagging on an event-by-event basis.

\subsection{Discussion}\label{subsec:discuss}
With this non-trivial LHC application detailed in Sect.~\ref{sec:applic}, we have shown that one can assert the polarisation fraction on an event-by-event basis using amplitude information by reverting to machine learning. The method is per-cent accurate and particularly versatile.
In this section we discuss limitations of the methods as well as possible extensions, generalisations, and further applications.

\paragraph{Validity of the method}
In the present example the training phase has been performed on events spanning a very inclusive sample.
The trained models have then be used on a reduced phase-space as for typical experimental analyses.
This ensures that the method is used in its region of validity.
We therefore recommend to always perform the training on a more inclusive phase-space than the one actually used in the analysis.
If this is not the case, it is not guaranteed that the method will still work as the network has not been trained (and thus validated) in the whole region considered at testing level.
While it is not excluded that the \NN can perform some extrapolation outside its training region, this has to be carefully verified.
In particular, using the extrapolation power of the \NN might require a different \NN and a potentially a dedicated study on out-of-support extrapolation problem into a problem of within-support generalization.

\change{
  While the specific application considered in this work concerns $\Pp\Pp\to\PZ+\Pj$ at the LHC,
  the general idea of training a NN with experimentally accessible kinematic information and
  squared polarised amplitudes can be applied to any single- or multi-boson process at colliders.
  We stress that in order to apply this strategy to another process, a new \NN has to be
  constructed, relying on the corresponding input features that depend on the experimental signature.
  For example, for processes with final-state neutrinos, whose momenta cannot be fully reconstructed,
  the input feature would be the missing transverse momentum instead of the complete momenta of the
  neutrinos.
}

\paragraph{Error propagation}

As formulated here, the method would provide a numerical value corresponding to the experimentally-extracted fraction that can be compared against theoretical predictions.
Nonetheless this extracted value has uncertainty of different sources: the accuracy of the theory prediction it relies on, the limited statistics of the training data set, and the experimental accuracy (both statistical and systematic) of the data.

Usually, the theoretical uncertainty on the prediction is assessed by means of scale variations of the factorisation and renormalisation scale. The envelop of the values of $r_\rL$ extracted for different scale combinations would then provide the theory uncertainty associated to $r_\rL$. This quantity being a ratio of squared amplitudes, we expect the correlated scale uncertainties to be rather small, owing to cancellations between the numerator (longitudinal matrix element) and the denominator (unpolarised).

The uncertainty related to the finite size of the training sample can be inferred by performing the training with different sample sizes or by performing error propagation in the \NN.
The same applies to the experimental error associated to the reconstructed event kinematics, and it can be estimated by repeating the method using pseudo-data.

It is important to consider that NNs are complex models, and training them using stochastic gradient descent over a non-convex landscape does not guarantee optimal parameter quality upon convergence. In contrast, linear and kernel models can be trained more efficiently due to their convex and low-dimensional loss landscapes, albeit resulting in simpler predictors. The non-linearity of NNs allows them to learn new and more effective representations of the data, a process known as \emph{feature learning} \cite{roberts2022principles}. This feature learning effect makes NNs more powerful but also presents challenges in their training process. In our approach, we have opted for wide NNs that strike a balance between linear and nonlinear models. While infinite-width NNs are equivalent to linear models and enjoy convex optimization landscapes \cite{jacot2018neural}, wide networks with finite width exhibit a slightly more challenging training landscape. Nevertheless, training wide NNs remains effective, with the difficulty of the landscape increasing as the depth-to-width ratio grows. Considering all these aspects related to the complexity behavior of ML models, we refrain from assigning any intrinsic uncertainty to the predicted output.

\paragraph{Model independence}

As already mentioned, the polarisation of weak bosons is a pseudo-observable and its extraction necessarily understands some degree of model dependence.
In the method we propose, the model dependence is encoded into the $r_\rL$ function.
In the present work, the SM is considered: it means that events are tagged according to SM expectations and the longitudinal fractions extracted should be compared against the one of theoretical predictions within the SM. 
Such a model dependence is impossible to avoid.
In fact, even a simplified version of $r_\rL$ relying only on the boson-decay matrix elements still depends on the polarisation fractions determined by the model-specific production mechanism.
However, since the method proposed in this work is model agnostic, the same procedure can be performed with more general models, \ie {simplified models} or {effective field theories}, or even with UV-finite theories.

\paragraph{Extension to higher orders}

In the present work, we have restricted our analysis to LO+PS accuracy.
Nonetheless, it is in principle possible to extend this to higher orders in perturbation theory, at least for QCD corrections to processes with leptonically decaying bosons.
If one can produce a sample of unweighted events at a given order in perturbation theory, the presented method can be extended.

Having unweighted events at fixed order implies having events with different jet multiplicity (depending on the order considered).
It means that for each multiplicity $i=0,1,...$, the exact $r_\rL^i$ can be computed using loop and/or tree amplitudes depending on the accuracy of the sample.
As ratios, they should actually be free of infrared singularities if QCD dependencies factorise from the polarisation effects.
As in the presented application, the \NN can learn a single approximate $\tilde r_\rL$ based on the experimental input available and in particular by feeding only the leading jet(s) in the transverse momentum.
Adding PS or further corrections can then be achieved as shown in the previous sections.

We stress that we have not explicitly tested this method with higher orders and therefore that if ones wants to use this proposal, it should be carefully checked first.
In particular, the main assumption here is that QCD corrections and polarisation effects factorise to a large extend (as PS and polarisation effects).
This implies that the inclusion of EW corrections would probably requires a more refined analysis given that they are known not to factorise.

\paragraph{Generalisation to other problems}

As highlighted several times, the key aspect of the method is to encode the whole physics problem in one single ratio (in the present application $r_\rL$) which can be approximately reconstructed using incomplete information thanks to \NN methods.
It therefore implies that the method can be applied to any physics problem that can be cast in this form.
The only requirement being that the key quantity is bounded as it is the case for ratios of amplitudes.
It also means that appropriate problems for this method are the extraction of a signal over a background which is very common in experimental particle physics.

\section{Conclusions}\label{sec:concl}

The polarisation of heavy gauge bosons encodes the intricate structure of the electroweak sector of the Standard Model.
The theoretical study and the experimental extraction of such pseudo-observables is thus of prime importance for the present and upcoming physics programme of the LHC.
It is therefore key to combine our theoretical understanding to make use of all the information available in experimental data in order to probe the structure of the Standard Model at the deepest.

In this work, we have designed an original method to extract polarisation fractions using the maximal information encoded in the amplitude thanks to the versatility of neural networks.
The key feature is that all information is encapsulated in a single number which can be computed on a event-by-event basis.
In particular, the neural network is able to construct a particularly good approximation of this quantity which can then be evaluated with incomplete information, namely the one available in experiments.
This number allows to assert whether an event is most likely longitudinally polarised or not.
In this way, all information \emph{i.e.}\ the fully differential information is exploited and not only the information contained in one or several observables as it is the case for other methods.
It also means that no fitting procedure is required.
Another advantage is that the theory dependence is clearly identified as it is only encoded in the amplitude.
Finally, the amplitude considered can be the one of arbitrary-general or -specific models of quantum field theory.

To illustrate the method, we have applied it to the extraction of the longitudinal polarisation of a $\PZ$ boson in the hadronic process $\Pp\Pp\to\PZ+\Pj$, in the leptonic decay channel at the LHC.
We have demonstrated that the idea is working with a per-cent accuracy by reverting to the sequential training of a neural network.
In particular, when being used in actual experimental analyses, the closure tests that we have presented here should be carried out to ensure the correctness of the results.

Finally, we point out that the method we have developed is very general.
It can therefore be applied to other problems and/or generalised.
In particular, the method seems to be particularly appropriate for the extraction of signals over irreducible or even reducible background.

\section*{Acknowledgements}

G.P.\ would like to thank Marco Zaro for useful discussion on technical aspects of \sloppy \madgraphbis, Rhorry Gauld and Silvia Zanoli for helpful suggestions regarding {\scshape{Pythia8}} settings.
M.I.\ is partially supported by the Istituto Nazionale di Alta Matematica ``Francesco Severi''.
M.G.\ is supported by CERN through CERN Quantum Technology Initiative.
M.P.\ acknowledges support by the German Research Foundation (DFG) through the Research Training Group RTG2044. G.P. was supported by the German Federal Ministry for Education and Research (BMBF, contract no.~05H21WWCAA) when this project was initiated.

\bibliographystyle{utphys.bst}
\bibliography{polarisation}
\end{document}